\documentclass[aps,pra,twocolumn,nofootinbib,superscriptaddress]{revtex4-2} 
	
	\usepackage{mathtools,amsmath}
	\usepackage{graphicx} 
	\usepackage{xcolor}
	\usepackage[breaklinks=true,colorlinks,citecolor=blue,linkcolor=blue,urlcolor=blue]{hyperref}
	\usepackage{physics}
	\usepackage{bbold}
	\usepackage{soul}
	\usepackage{bm}
	\usepackage{braket}
	\usepackage{wrapfig}
    \usepackage{nicefrac}
    \usepackage[normalem]{ulem}
    \usepackage{array}
    \usepackage{multirow}
    \usepackage{enumitem}
    \usepackage{tabularray}
    \usepackage{changes}
    
    \usepackage{pifont}
    \newcommand{\cmark}{\text{\ding{51}}}
    \newcommand{\xmark}{\text{\ding{55}}}
    
    \usepackage{svg}

	\newcommand{\D}{{\rm d}}
	\newcommand{\C}{\mathcal{C}}

\begin{document}
\title{
Optimal Thermometers with Spin Networks 
 } 
\begin{abstract}
The heat capacity $\C$ of a given probe is a fundamental quantity that determines, among other properties, the maximum precision in temperature estimation.  
In turn,  $\C$ is limited by a quadratic scaling with the number of constituents of the probe, which provides a fundamental limit in quantum thermometry. 
Achieving this fundamental bound with realistic probes, i.e. experimentally amenable, remains an open problem. 
In this work, we tackle the problem of engineering optimal thermometers by using networks of spins. Restricting ourselves to two-body interactions, we derive general properties of the optimal configurations and exploit machine-learning techniques to find the optimal couplings.
This leads to simple architectures, which we show  analytically to approximate the theoretical maximal value of $\C$ and maintain the optimal scaling for short- and long-range interactions. Our models can be encoded in currently available quantum annealers, and find application in other tasks requiring Hamiltonian engineering,  ranging from quantum heat engines to adiabatic Grover’s search.
\end{abstract}



\author{Paolo Abiuso}
\affiliation{ICFO – Institut de Ci\`encies Foto\`niques, The Barcelona Institute of Science and Technology, 08860 Castelldefels (Barcelona), Spain}
\affiliation{Department of Applied Physics, University of Geneva, 1211 Geneva, Switzerland}
\affiliation{Institute for Quantum Optics and Quantum Information - IQOQI Vienna,
Austrian Academy of Sciences, Boltzmanngasse 3, A-1090 Vienna, Austria}

\author{Paolo Andrea Erdman}
\affiliation{Freie Universit{\" a}t Berlin, Department of Mathematics and Computer Science, Arnimallee 6, 14195 Berlin, Germany}

\author{Michael Ronen}
\affiliation{Deparment of  Physics, University of Konstanz, D-78457 Konstanz, Germany}
\affiliation{Department of Applied Physics, University of Geneva, 1211 Geneva, Switzerland}

\author{Frank No{\'e}}
\affiliation{Microsoft Research AI4Science, Karl-Liebknecht Str. 32, 10178 Berlin, Germany}
\affiliation{Freie Universit{\" a}t Berlin, Department of Mathematics and Computer Science, Arnimallee 6, 14195 Berlin, Germany}
\affiliation{Freie Universit{\" a}t Berlin, Department of Physics, Arnimallee 6, 14195 Berlin, Germany}
\affiliation{Rice University, Department of Chemistry, Houston, TX 77005, USA}

\author{Géraldine Haack}
\affiliation{Department of Applied Physics, University of Geneva, 1211 Geneva, Switzerland}

\author{Mart\' i Perarnau-Llobet}
\affiliation{Department of Applied Physics, University of Geneva, 1211 Geneva, Switzerland}

\maketitle




\section{Introduction}

Our ability to measure temperature in quantum systems is currently being pushed to new regimes~\cite{Giazotto2006,Yue2012,pasquale2018quantum,mehboudi2019thermometry}. 
At the experimental level, ultraprecise temperature measurements of  gases at the lowest temperatures in the universe are  possible~\cite{Bloch2008,Onofrio_2016}, and new methods for  thermometry with probes of atomic size are being developed. Relevant examples include nanodiamonds acting as thermometers of living cells~\cite{kucsko2013nanometre,fujiwara2020real},  nanoscale electron calorimeters based on the absorption of single quanta of energy~\cite{Gasparinetti2015,halbertal2016nanoscale,karimi2020reaching},  and  single-atom thermometry probes~\cite{Hohmann2016,Bouton2020,Adam2022}. 
At the theoretical level, progress has been made in the understanding of ultraprecise thermometry via quantum probes in equilibrium~\cite{Hovhannisyan2018,Potts2019,Jorgensen2020,Mukherjee_2019,Glatthard2022, Correa2017,Mehboudi2019,Planella2022,Khan2022}
and out-of-equilibrium states~\cite{Brunelli_2011, Brunelli_2012, Jevtic_2015, Guo_2015, Pasquale_2017,Hofer2017, Cavina_2018, Mancino_2020,Mitchison2020,Hovhannisyan2021,Zhang2022}.
Crucially, the energy structure of optimal thermometers has been revealed~\cite{correa2015individual,Campbell2018,mok2021optimal,sekatski2021optimal}, suggesting that the precision can grow quadratically with the number of constituents~\cite{mehboudi2022fundamental}. There is however still a gap between such theoretical bounds  and state-of-the art experimental implementations, which is crucial to address to exploit the full potential of quantum thermometry. 

Due to its generality and practical relevance, we consider in this work equilibrium thermometry~\cite{mehboudi2019thermometry}. In this case, the probe is assumed to be well described by a thermal state at the temperature $T$ that is being estimated. 
Then,  the error $\Delta T$ of any measurement on the probe is bounded by~\cite{Old_PRE,Paris_2015}:
\begin{align}
    \frac{\langle(\Delta T)^2\rangle}{T^2} \geq \frac{1}{ \nu\mathcal{C}}\;,
    \label{Eq:precision}
\end{align}
where $ \mathcal{C}$ is the heat capacity of the probe, and $\nu$ the number of repetitions of the experiment -- see  Sec.~\ref{Sec:EquilibriumThermometry} for a precise definition of the quantities involved. 
Intuitively speaking, a high heat capacity ensures that the energy of the probe highly varies with $T$, thus enabling the detection of small temperature variations.  

An optimal probe for  equilibrium  thermometry is hence the one with the highest heat capacity. 
The ultimate limits to this problem were set by  Correa et al. in Ref.~\cite{correa2015individual} by finding the maximum $\C$ given an arbitrary Hamiltonian of dimension $D$.  
The  spectrum of such an optimal probe consists in an effective two-level system, with a single ground state and an exponential degeneracy of the excited level. The resulting optimal heat capacity reads $\mathcal{C}^{\rm opt}\approx (\ln D)^2/4$. If we consider that the probe consists of $N$ bodies of dimension $d$ (hence $D=d^N$), then $\mathcal{C}^{\rm opt}$ becomes~\cite{correa2015individual,mehboudi2022fundamental}:
\begin{align}
   \mathcal{C}^{\rm opt}\approx \frac{N^2 (\ln d)^2}{4}.
    \label{Eq:UltimateLimit}
\end{align}
This expression shows a quadratic scaling with the number of constituents~$N$, to be confronted with the typical extensive behaviour of the heat capacity (i.e. linear in~$N$). This quadratic scaling is reminiscent of the well-known Heisenberg limit in quantum metrology~\cite{giovannetti2006quantum}, although it should be realised that the advantage here arises due to the interacting nature of the probe's Hamiltonian, and not from the presence of entanglement in the probe. Reference~\cite{mok2021optimal} provides a specific $N$-spin interacting Hamiltonian that can saturate \eqref{Eq:UltimateLimit}, which however requires  $N$-body interactions.  A natural question therefore arises: \begin{enumerate}
\item[{\bf Q:}] \emph{Can we reach the ultimate limit \eqref{Eq:UltimateLimit} via realistic Hamiltonians, i.e.   featuring two-body and local interactions? } 
\end{enumerate}

A natural approach to address {\bf Q} is to consider probes at the verge of a thermal phase transition, where the heat capacity can scale superextensively with~$N$~\cite{zanardi2007bures,zanardi2008quantum,salvatori2014quantum,Mehboudi_2015,Salado_Mej2021spectroscopy,Aybar2022criticalquantum}. 
Previous studies  with  spin systems close to criticality exemplify the potential of phase transitions for thermometry~\cite{salvatori2014quantum,Mehboudi_2015,Aybar2022criticalquantum} but  do not come close to the ultimate limit~\eqref{Eq:UltimateLimit}. 
For small values of $N$, proposals for optimal probes  have also been considered with spin chains~\cite{Guo2015,mok2021optimal} or interacting fermionic systems~\cite{Marcin2018}.  
Yet, despite promising progress, none of the above approaches leads to a general answer to~{\bf Q} and hence to the possibility of approaching a quadratic precision in quantum thermometry. 

 

To address~{\bf Q}, we consider as a platform a generic system of spins with two-body interactions, such as those currently programmable in quantum annealers.  
Their open system dynamics is starting to be studied~\cite{benedetti2016estimation,marshall2019power,buffoni2020thermodynamics,izquierdo2021testing,morrell2022signatures}, and they represent flexible physical devices with a high degree of control. 
More specifically, we consider a  Hamiltonian of the form: 
\begin{align}
\label{eq:classical_H_generic}
    H=\sum_i^N h_i \sigma^{z}_i + \sum_{i < j}^N J_{ij} \sigma^{z}_i \sigma^{z}_j \;,
\end{align}
where $\sigma_i^z=\pm 1$ is the $i$-th classical spin of the system~\footnote{In the Appendix, we consider also the case of fully quantum-mechanical  spin Hamiltonians. However, the numerical optimization, for $N$ up to $9$, suggests that no advantage is given by considering the most general two-body Hamiltonian with arbitrary   off-diagonal interactions involving also $\sigma^x$ and $\sigma^y$ terms  (see App.~\ref{app:quantum} for details).}.
We then maximise~$\mathcal{C}$ over all control parameters $h_i$ and $J_{ij}$ (with different constraints on their locality and strength).   To tackle the exponential complexity of this task, 
we use advanced numerical techniques, commonly employed in the Machine-Learning community, to discover ansatz for the form of optimal probes. 
We then combine these numerical ansatz with physical insights to analytically prove 
that ~$\C$ can display the quadratic scaling of Eq.~\eqref{Eq:UltimateLimit}, with a slightly worse prefactor that depends on the locality of the Hamiltonian~\eqref{eq:classical_H_generic}, thus answering affirmatively~{\bf Q}.
These results add on recent applications of Machine-Learning based techniques in the field of quantum thermodynamics~\cite{hernandez2021,sgroi2021,khait2021,ashida2021,erdman2022,erdman2022_blackbox,erdman2022_pareto,luiz2022}, as well as in other  domains, including protein folding~\cite{jumper2021}, many-body problems~\cite{carleo2017, noe2019, hermann2020}, geosciences \cite{bergen2019},~algorithm discovery \cite{fawzi2022}.

\begin{figure}[t]
  \includegraphics[width=\columnwidth]{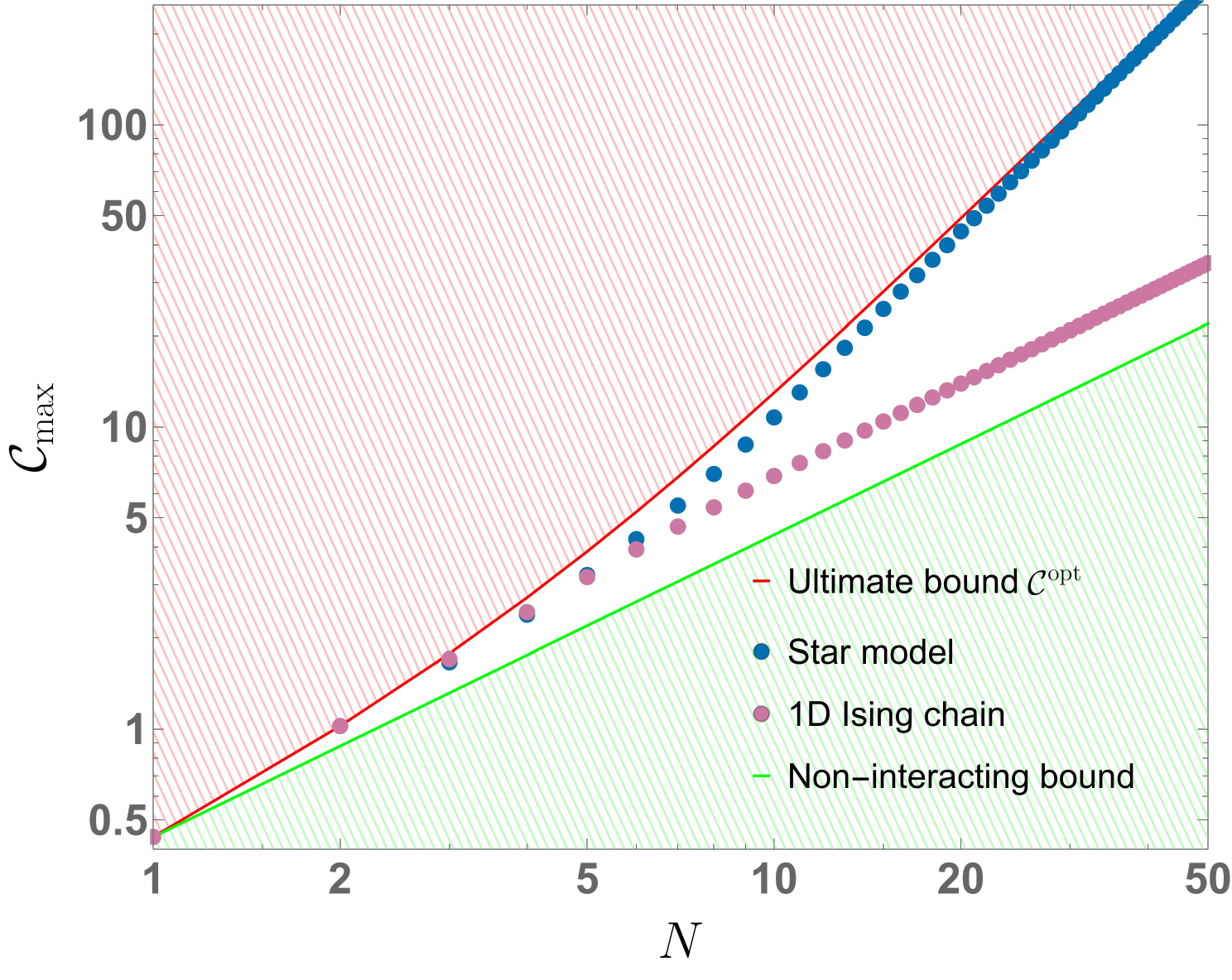}
  \caption{
  Maximum heat capacity $\C_{\rm max}$ of spin-based thermometers.
  The red line corresponds to the mathematical bound $\mathcal{C}^{\rm opt}$~\eqref{Eq:UltimateLimit} on any system with dimension $D=2^N$ (for a formal definition, see Eq.~\eqref{eq:Copt_def}), which shows a quadratic $\propto N^2$ scaling in terms of the number $N$ of total spins employed.
  Our optimal spin-network architecture, the ``Star model", provides the highest heat capacity for Hamiltonians of the form~\eqref{eq:classical_H_generic} when $N\geq 6$, and can reach the mathematical bound $\mathcal{C}^{\rm opt}$~\eqref{Eq:UltimateLimit} in the large $N$ limit. 
  This is to be compared with the extensive $\propto N$ scaling of standard models, such as the 1D Ising chain. 
  The green line delimits the region accessible with non-interacting spins, and simply corresponds to $\sim 0.44 N$, $0.44$ being the maximum heat capacity of a single spin. 
  } 
  \label{fig:smallscalingplot}
\end{figure}

In Fig.~\ref{fig:smallscalingplot}, we illustrate the type of results obtained.
The heat capacity of any spins system is upper-bounded by the fundamental bound $\C^{\rm opt}$ (red line, and Eq.~\eqref{Eq:UltimateLimit} with $d=2$), however
the maximum heat capacity obtainable with $N$ non interacting spins simply corresponds to $N$ times the maximum heat capacity of a single spin (green line).
The use of interactions can enhance  $\C$. 
Nevertheless, standard interacting spin-networks such as the 1D Ising model in the Figure (purple dots) show an extensive scaling of~$\C_{\rm max}$ in the limit of large $N$, hence losing their advantage. 
In contrast, we find optimal spin-network architectures~\eqref{eq:classical_H_generic} that approximate $\C^{\rm opt}$ for all $N$, see e.g. the \emph{Star model} as an example of the architectures discussed in the next sections (blue dots in Fig.~\ref{fig:smallscalingplot}).

The rest of the paper is structured as follows. In Sec.~\ref{Sec:EquilibriumThermometry} we review equilibrium thermometry, 
and we analyse the fundamental properties of the optimal energy spectra for the maximization of $\C$.
In Sec.~\ref{sec:optimal_spin_probes} we move to the case of physically realistic probes~\eqref{eq:classical_H_generic}, we present the derivation and analysis of our optimal thermometer models, and
then discuss their implementation and properties. 
In Sec.~\ref{sec:other_models} we describe other relevant models which we use for performance comparison.
Finally in Sec.~\ref{sec:conclusions} we conclude and discuss future directions and applications of this work. 
The Appendix contains details of the numerical methods employed, technical analytical derivations, and complementary analysis of our results.
The code written to perform the machine-learning based optimization will be available online when published (see the ``Code availability'' section).

\section{Equilibrium thermometry and properties of optimal spectra}
\label{Sec:EquilibriumThermometry}

Let us consider a sample at some unknown temperature $T$, corresponding to the inverse temperature $\beta=T^{-1}$ (herafter we set $k_B = 1$ for simplicity). To assess $\beta$, we let the sample weakly interact with a probe described by its Hamiltonian $H$. After a sufficiently long time, it is assumed that the probe will reach a Gibbs state, fully determined by $H$ and $\beta$: 
\begin{align}
\rho_\beta(H):=\frac{e^{-\beta H}}{\Tr[e^{-\beta H}]}\;.
\label{eq:GibbsState}
\end{align}
By measuring the energy of $\rho_\beta(H)$, it is possible to infer $\beta$ (hence the temperature $T$). Let us note that projective energy measurements were shown to be optimal for temperature estimation
~\cite{Paris_2015,correa2015individual}. 
In particular, the Cramer-Rao bound~\cite{cramer2016mathematical} specific to the case of temperature estimation~\cite{pasquale2018quantum,mehboudi2019thermometry} can be exploited to estimate minimal error $\Delta T$.
More precisely, for a number $\nu$ of identically and independently  distributed (i.i.d.) repetitions of the experiment, $\Delta T$ has a mean square value that is bounded by Eq.~\eqref{Eq:precision}, that is  $\frac{\langle(\Delta T)^2\rangle}{T^2} \geq (\nu\mathcal{C})^{-1}$.
It is therefore clear that the maximum precision one can get in estimating the temperature $T$ by measuring the energy of the probe at equilibrium with the sample 
is determined by the heat capacity $\C$.  
This one is formally defined as the variation in mean energy of the probe per temperature change unit, i.e.
\begin{align}
\label{eq:C_def}
    \mathcal{C}(H,\beta): =\frac{\D}{\D T}\Tr[H\rho_\beta(H)]
    =
    -\beta^2\frac{\D}{\D\beta}\Tr[H\rho_\beta(H)]\;.
\end{align}
In terms of the eigensystem $\{E_i, \ket{E_i}\}$ of the probe's Hamiltonian $H$, the state populations of $\rho_\beta(H)$ read $p_i\equiv Z_\beta^{-1} e^{-\beta E_i}$, with $Z_\beta=\sum_i e^{-\beta E_i}$ and $\rho_\beta=\sum_i p_i \ket{E_i}\bra{E_i}$.
In the energy eigenbasis of the probe, it is easy to verify from Eq.~\eqref{eq:C_def} that the heat capacity is proportional to the energy variance of the Gibbs state:
\begin{align}
    \C (H,\beta) =& \beta^2\Delta_\beta^2 H\;, \\
\label{eq:varianceDiag}
    \Delta_\beta^2 H =& \sum_{i=1}^D p_i E_i^2 - \left(\sum_{i=1}^D p_i E_i\right)^2,
\end{align}
where $D$ is the dimension of the Hilbert space.
Such expression clarifies that the heat capacity only depends on the spectrum of the Hamiltonian $H$ and inverse temperature $\beta$. It is important for the following to note the scale invariance $\C(\lambda H,\lambda^{-1}\beta)=\C(H,\beta)$, with $\lambda \in \mathbb{R}$. This allows us to express all energies in units of $\beta$, and to simply refer to the heat capacity as a function of a adimensional Hamiltonian $\tilde{H}:=\beta H$, as $\C(\tilde{H},1)=\C(H,\beta)$.
In the following, we will omit the tilde and simply use adimensional units, writing $\C(H):=\C(H,1)$. We also emphasize that a global energy shift does not affect neither the Gibbs state nor the heat capacity as $\C(H)=\C(H+c\mathbb{1}), c \in \mathbb{R}$.


\subsection{Optimal spectrum for equilibrium thermometry}
\label{sec:optimal_spectra}
From Eq.~\eqref{Eq:precision}, we see that an optimal probe for thermometry is the one with maximum heat capacity $\C$. The maximization of $\mathcal{C}$ of a generic $D$-dimensional system at thermal equilibrium has been carried out in~\cite{correa2015individual} assuming full-control on the Hamiltonian and its spectrum,
\begin{align}
    \C^{\rm opt}(D):= \max_{H | {\rm dim }H=D} \C (H).
    \label{eq:Copt_def}
\end{align}
The resulting optimal spectrum consists of a single ground state and a $(D-1)$-degenerate excited state,
that is
\begin{align}
\label{eq:deg_Ham}
    H_{\rm deg}=0\ketbra{0}{0}+\sum_{i=1}^{D-1} E\ketbra{i}{i}\;,
\end{align}
with an \textit{optimal gap} $E=x$ in temperature units that satisfies the transcendental equation $e^x=(D-1)(x+2)/(x-2)$.
The corresponding heat capacity is $\mathcal{C}^{\rm opt}(D)=x^2e^x (D-1)/(D-1+e^x)^2$~\cite{correa2015individual}.
This expression gives in the asymptotic regime of large probes ($D\rightarrow\infty$) $x \simeq \ln D$, hence $\mathcal{C}^{\rm opt}(D) \simeq (\ln  D)^2/4$.  For a probe made up of $N$ constituents, each with local dimension $d$, so that $D=d^N$, we recover the Heinsenberg-like scaling in Eq.~\eqref{Eq:UltimateLimit}.



 \subsection{Properties of optimal spectra} 
 \label{sec:spectra_properties}
In order to understand the origin of the desired scaling $\C \propto N^2$, we now discuss the relevant features of the spectrum Eq.~\eqref{eq:deg_Ham} in its optimal configuration, as well as possible perturbations of it. This will be relevant for cases in which a physical realization (e.g. using Eq.~\eqref{eq:classical_H_generic}) can approximate Eq.~\eqref{eq:deg_Ham}, but not exactly. Specifically, we prove that a large class of spectra can exhibit the Heisenberg-like scaling $\propto N^2$ of the heat capacity, when the following 3 properties are satisfied: 

{\bf P1: exponential degeneracy.} The spectrum has a two-level structure with single ground state, and a first excited level that is exponentially degenerate (in $N$), with a gap that can be tuned.

{\bf P2: bandwidth tolerance.} The engineering of the effective two-level  spectrum, and in particular of the bandwidth of the first excited level, can tolerate a relative precision of $\mathcal{O}(1/N)$.

{\bf P3: tolerance to additional energy levels.} The presence of other energy levels does not necessarily deteriorate the maximal value of $\C$ and its scaling. In particular: \emph{i)} high energy levels (i.e. above the first excited level) do not decrease the maximal heat capacity, while \emph{ii)} energy levels below the first excited have an exponentially small contribution to the heat capacity, provided that their total degeneracy is (at most) polynomial in $N$, and their gap to the ground level increases (at least) linearly in $N$. 

%
%

\begin{figure}
    \centering
    \includegraphics[width=\columnwidth]{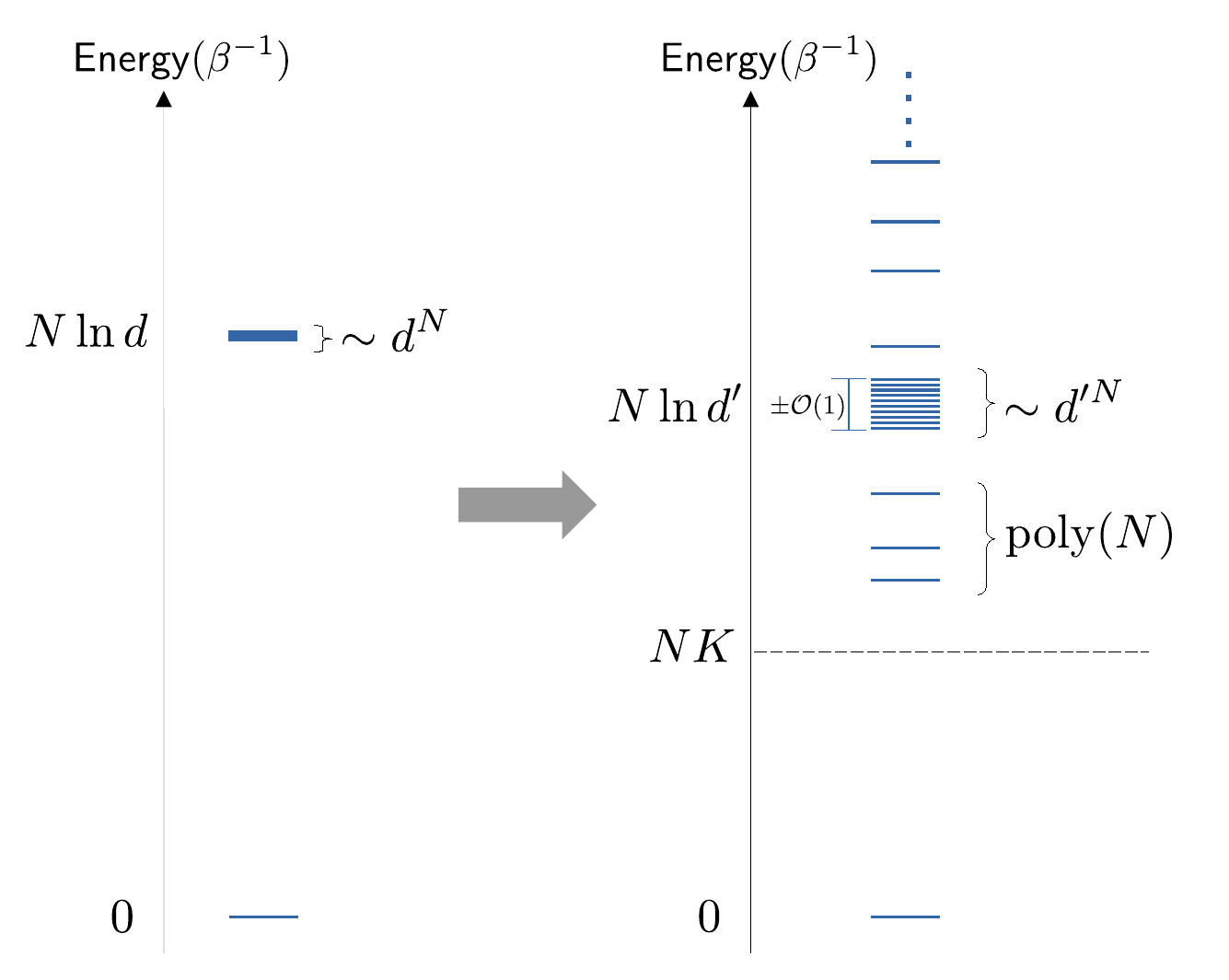}
    \caption{(Left) The idealized model $H_{\rm deg}$~\eqref{eq:deg_Ham}. (Right) we prove that any Hamiltonian featuring a spectrum of the form respecting properties {\bf P1-P2-P3} (see details in text) can exhibit a $\propto N^2$ scaling of the maximal heat capacity. 
    }
    \label{fig:levels_properties}
\end{figure}

A schematic representation of the class of spectra satisfying the above three properties is given in Fig.~\ref{fig:levels_properties}. We now  provide an intuitive understanding of these properties. 

The importance of the exponential degeneracy of the first excited state ({\bf P1}) can be appreciated from the degenerate model Eq.~\eqref{eq:deg_Ham} and its corresponding ground state probability for the Gibbs state (in units of $\beta$), $p_0=(1+(D-1)e^{- E})^{-1}$, which can be expressed as
\begin{align}
    p_0=\left(1+{\rm Exp}[\ln{(D-1)}-E]\right)^{-1}
\end{align}
For small energy gaps $E$ and large $D$, the value of $p_0$ is $\sim 0$, meaning that in the thermal state, almost all the population is spread evenly in the degenerate excited subspace. When the gap reaches $E\sim (\ln (D-1)) $, $p_0=\frac{1}{2}$, while for larger values 
it 
increases to $\sim 1$, and 
the excited levels become empty. The width of this transition is of order $\sim \mathcal{O}(1)$, and it is the point where the system experiences the peak in heat capacity; in fact, for smaller (larger) values of $E$, the energy variance is suppressed exponentially, given that the whole population collapses to the excited subspace (ground state). At the peak of the heat capacity, approximately half of the population is in the ground state, and half is spread in the degenerate level. If the degeneracy $(D-1)= d^N-1$ is exponential in $N$, the optimal gap is linear in $N$, and the resulting energy variance Eq.~\eqref{eq:varianceDiag} scales quadratically. 
According to this observation, the exponential (in $N$) degeneracy of the first excited level is the first main ingredient for a system to exhibiting such quadratic scaling of the heat capacity. Furthermore, we 
notice that at a formal level, the same scaling is obtained whenever $D\propto d'^N$ for some  $d'>1$, which leads to {\bf P1}.
Notice however that any physical implementation of such a conceivably highly fine-tuned two-level probe will be susceptible to noise. The resulting deviation will cause a broadening of the ideally degenerate excited level into a band. 
In App.~\ref{app:noise_tol}, we prove that the optimal scaling of $\C$ is preserved as long as the error in the energy gap between the ground state and the first excited state (including the broadening of level band) is of order $\mathcal{O}(1)$. This is to be contrasted with an optimal gap $E\propto N$ that scales linearly, thus requiring  a relative precision of $1/N$ in the engineering of the energy levels ({\bf P2}).

Finally, it is possible to show that (the quadratic scaling of the heat capacity?)  is preserved even in the presence of additional ``undesired" energy levels, provided that property {\bf P3} is satisfied. 

More precisely, consider two Hamiltonians, $H_1$ with dimension $1+k_1$, and $H_2$ with dimension $1+k_1+k_2$. $H_1$ has 1 ground state and a $k_1$-degenerate excited state, $H_1=0\ketbra{0}{0}+\sum_{i=1}^{k_1}  E \ketbra{i}{i}$,
while $H_2$ has the same spectrum and additional $k_2$ excited states above,
$H_2 =H_1+ \sum_{\alpha=k_1+1}^{k_1+k_2} E_\alpha \ketbra{\alpha}{\alpha}$,
with $0\leq E\leq E_\alpha\; \forall\alpha$.
Assuming control over the first excited gap $ E$, we prove in App.~\ref{app:Lemma_proof}  that the  maximal achievable heat capacity with $H_2$ is always larger than the maximal achievable heat capacity with $H_1$,
\begin{align}
    \max_{E}\C(H_1)\leq \max_{E\leq E_\alpha} \C(H_2)\;.
    \label{eq:prop2}
\end{align}
This property guarantees that additional excess levels above the $k_1$-degeneracy of $H_1$ can only increase the maximal heat capacity. As a consequence, as the system size grows, any model featuring an exponential degeneracy of the first excited level and a tunable gap will show the desired Heisenberg-like scaling of the heat capacity. The control over $ E$, while keeping $E_\alpha\geq  E\; \forall\alpha$, can easily be obtained, for example by rescaling all the parameters of $H_1$ or $H_2$ globally.
For what concerns additional levels \emph{below} the first excited, it is enough to notice that if their total number is of order $\mathcal{O}(N^k)$ for finite $k$, and their gap from the ground state energy is bounded between $NK$ and $N\ln d'$ ($0<K<\ln d'$), their total contribution to the variance~\eqref{eq:varianceDiag} scales as $\mathcal{O}(N^{k+2} \exp[-\beta NK])$, and is therefore suppressed for large $N$.

\section{Optimal spin-network thermometers}
\label{sec:optimal_spin_probes}


We recall that, without any restriction on the possible interactions among the $N$ spins, it is possible to generate the Hamiltonian Eq.~\eqref{eq:deg_Ham} and to saturate the theoretical maximum value $\C^{\rm opt}$ of the heat capacity (see e.g.~\cite{mok2021optimal}, where the authors make use of arbitrary $N$-body interactions). The question {\bf Q}  we address in this work is 
whether it is possible to achieve the optimal scaling $\mathcal{C} \propto N^2$ if we restrict ourselves to  physically motivated $2$-body Hamiltonians given by Eq.~\eqref{eq:classical_H_generic}.
In such spin-systems, we have 
$ D=2^N $,
where $N$ is the total number of spins, thus the ultimate limit Eq.~\eqref{Eq:UltimateLimit} reads
\begin{align}
\label{eq:optimal_scaling}
\mathcal{C}^{\rm opt}(2^N) \simeq \frac{N^2 (\ln 2)^2}{4}\;, \quad
\beta E\simeq {N}{\ln 2}\;,
\end{align}
for large $N$.
Below, we demonstrate that the answer to our main question is positive. We show that it is possible to design a thermal probe (``Star model", Sec.~\ref{sec:star_model}) consisting of $N$ interacting spins with two-body interactions that approximates the maximum value $\mathcal{C}^{\rm opt}$ of the thermal sensitivity Eq.~\eqref{eq:optimal_scaling}. 
We further prove that a thermal probe (``Star-chain model'', Sec.~\ref{sec:star_chain_model}) with two-body and local interactions can be designed with a heat capacity exhibiting the same scaling as Eq.~\eqref{eq:optimal_scaling} with a prefactor that can be made arbitrarily close to the Star model. Moreover, in Sec.~\ref{sec:chimera_embed} we show that the Star-chain model can be realized on currently available quantum annealers. Finally, in Sec.~\ref{sec:scaling-constraints} we analyze the scaling of the Hamiltonian parameters in these configurations, and the effect of  constraints on the absolute value of the parameters.

\subsection{Star model}
\label{sec:star_model}

We now search for thermal probes, consisting of spin networks with two body interactions, that maximize the heat capacity. We 
maximized $\C(H)$ over the parameters $h_i$ and $J_{ij}$ employing Eq.~(\ref{eq:varianceDiag}) and constraining $H$ to be of the form (\ref{eq:classical_H_generic}). Notice that such problem is numerically hard due to (i) its nonconvexity, (ii) the number of optimization parameters that scales quadratically in $N$, and (iii) the number of spin configurations that scales exponentially. As such, first attempts based on simpler techniques such or gradient descent with momentum~\cite{qian1999} estimating the gradients with finite-differences, and the gradient-free covariance matrix adaptation evolution strategy~\cite{hansen2003}, 
would get stuck in sub-optimal local maxima with a substantially lower $\C$, not exhibiting the Heisenberg-like scaling. 
We thus decided to use tools commonly employed in Machine Learning, i.e. we implemented the optimization in PyTorch that allows us to compute the exact gradients of the negative heat capacity using backpropagation~\cite{goodfellow2016}, and we used the Adam optimizer~\cite{kingma2014}
(see App.~\ref{sec:star_optimization} for details).

\begin{figure}[!tb]
	\centering
	\includegraphics[width=0.99\columnwidth]{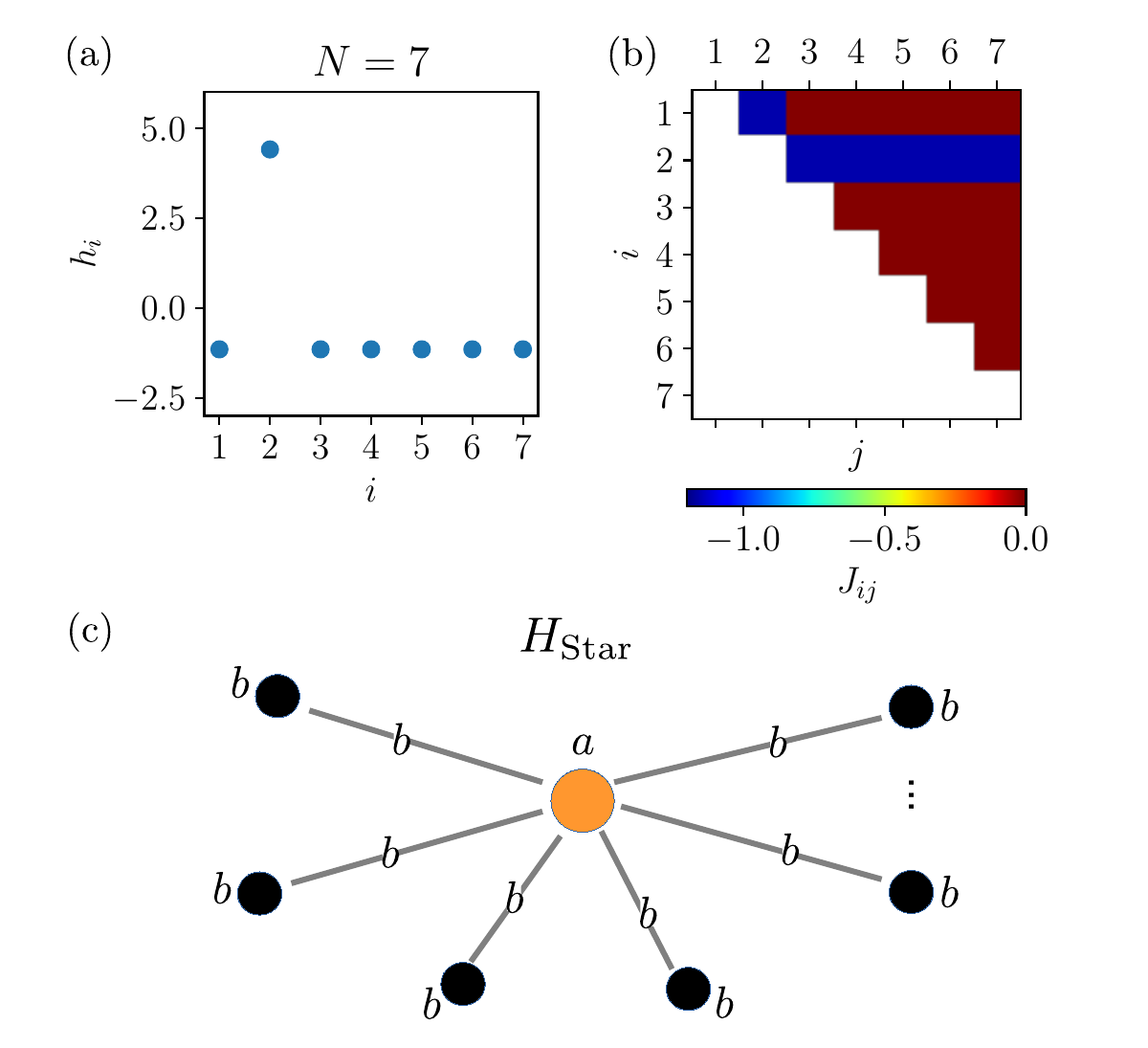}\\
\caption{(a,b): Machine learned Hamiltonian parameters~\eqref{eq:classical_H_generic},
for $T=1$ and $N=7$. (a) shows the local field $h_i$ as a function of the spin index, while (b) shows the $J_{ij}$ parameters (color) as a function of the site indices $i$ and $j$. The resulting model that emerges, sketched in (c), is $H_{\rm Star}$~\eqref{eq:H_star1}. It consists of a single central spin (corresponding to spin nr.~2 in (a,b), and orange circle in (c)) with a different local magnetic field and that interacts (gray lines) with all the other $N-1$ spins homogeneously (black circles) resulting in a Star-shaped connectivity. 
}
	\label{fig:star_scheme}
\end{figure}

\begin{figure}
    \centering
    \includegraphics[width=0.45\textwidth]{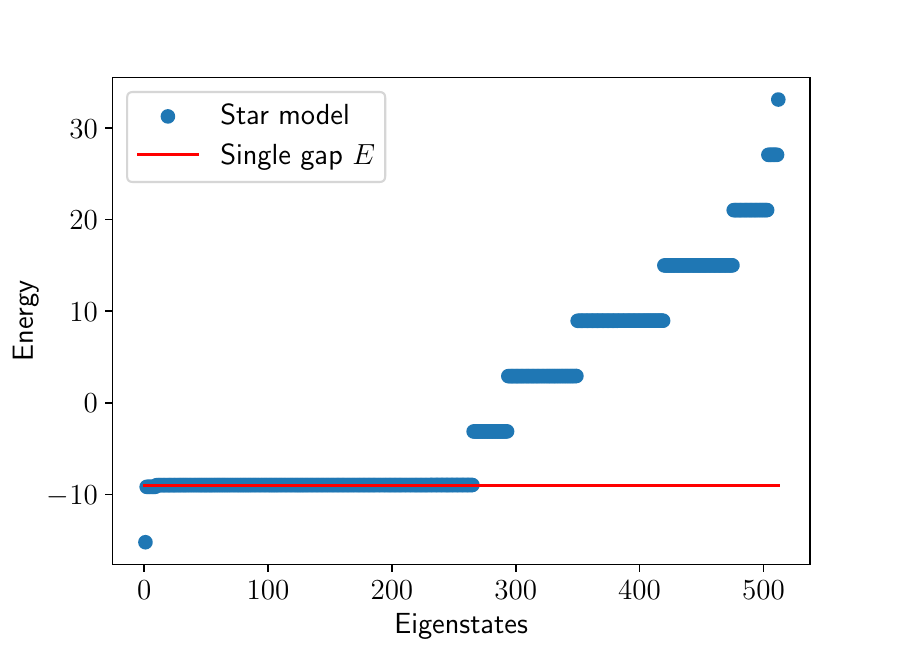}
    \caption{Spectrum of the Star model~\eqref{eq:H_star1} (for $N=9$). $2^{N-1}$ eigenvalues form a binomial spectrum~\eqref{eq:spectrum_star_binom}, while the other  $2^{N-1}$ eigenvalues are completely degenerate~\eqref{eq:spectrum_star_flat}. The gap between the ground energy and the exponentially degenerate level approximately coincides with the optimal gap of the ideal spectrum $H_{\rm deg}$~\eqref{eq:deg_Ham} for $D=2^{N-1}$~\cite{correa2015individual} (red line), as expected from the discussion in Sec.~\ref{sec:star_model}.}
    \label{fig:star_spec}
\end{figure}

After repeating the optimization for different total numbers of spins $N$, a recurrent pattern emerges (cf. App.~\ref{sec:star_optimization} and Fig.~\ref{fig:star_scheme}), 
corresponding to a ``Star model'' Hamiltonian of the form
\begin{align}
\label{eq:H_star1}
	H_{\text{Star}[N]}(a,b) &:= a \,\sigma_1^z + b \sum_{i=2}^N \sigma_i^z  \left(\mathbb{1} + \sigma_1^z\right)  \;,
\end{align}
with $a, b \in \mathbb{R}$, corresponding to a single spin ($\sigma_1^z$) that is coupled uniformly to all the other ones. A representation of this Star model is  shown in Fig.~\ref{fig:star_scheme}. 
The resulting spectrum has $2$ main classes of eigenstates. The first class consists of 
$\binom{N-1}{k}$-degenerate evenly spaced states with energy
\begin{equation}
	E_k = a + 2b(k-(N-1-k)),\quad \text{for }k=0,\dots,N-1\;,
	\label{eq:spectrum_star_binom}
\end{equation}
corresponding to the first spin being up, i.e. $\sigma_1^{z}=1$\;, and $k$ spins up among the remaining $N-1$ ones. The second class consists of the first spin being down $\sigma^z_1=-1$. In this the second term in~\eqref{eq:H_star1} becomes null, independently of the value of all the other spins $i=2,\dots,N$, and we get a
 $2^{N-1}$-degenerate excited state with energy 
\begin{equation}
	E_\text{deg} = -a\;.
	\label{eq:spectrum_star_flat}
\end{equation}
That is, thanks to the simple topology and choice of the couplings in Eq.~\eqref{eq:H_star1}, the first spin $\sigma^z_1$ acts as an ``on-off'' switch for the effective magnetic field on the remaining spins, generating an exponential degeneracy of the $E_{\rm deg}$ level. 
The partition function of the Star model can be solved analytically, being the sum of the two partition functions corresponding to $\sigma_1^z=\pm 1$, i.e.
\begin{align}
    Z_{\rm Star}=\left(e^{2\beta b}e^{-\beta a}+e^{-2\beta b}\right)^{N-1}+2^{N-1}e^{\beta a}\;.
\end{align}
This expression can be used to efficiently compute all the relevant thermodynamic quantities of the model (cf. App.~\ref{app:analytics_star}). Moreover, it is easy to see that by choosing
$b>0$ and $b(N-3)\leq a < b(N-1)$, one ensures
\begin{align}
    E_0 < E_{\rm deg}\leq E_k \quad \text{for }  k=1,\dots, N-1\;,
\end{align}
corresponding to a single ground state, and $2^{N-1}$-degenerate first excited level. By saturating $b(N-3)=a$, one gets $E_{\rm deg}=E_1$, corresponding to a $2^{N-1}+N-1$ degeneracy for the first excited state~\footnote{We conjecture that $2^{N-1}+N-1$ is the  maximal achievable degeneracy of the first excited level, in Hamiltonians of the form Eq.~\eqref{eq:classical_H_generic} with a single ground state.} (for a visual representation, see Fig.~\ref{fig:star_spec}).
Notice that property~{\bf{P3}}~\eqref{eq:prop2} ensures that such a model can achieve at least the heat capacity $\C^{\rm opt}(2^{N-1})$, that is
\begin{align}
    \mathcal{C}^{\rm opt}(2^{N-1})\leq \C_{\rm max}^{{\rm Star}[N]} \leq \mathcal{C}^{\rm opt}(2^N)\;.
\end{align}
In the asymptotic limit, we get
\begin{align}
\label{eq:optimal_star_scaling}
    \C_{\rm max}^{{\rm Star}[N]}\gtrsim \frac{(N-1)^2(\ln 2)^2}{4}\,,
\end{align}
which becomes indistinguishable from the theoretical bound $\C^{\rm opt}(2^N)$, see Eq.~\eqref{eq:optimal_scaling} and shown in Fig.~\ref{fig:smallscalingplot} and Fig.~\ref{fig:bigscalingplot} below. In App.~\ref{app:opt_values_tables}, we provide a table with the optimal values of the Hamiltonian parameters $a,b$ (see Eq.~\eqref{eq:H_star1}) and the corresponding value of $\C^{\rm Star[N]}_{\rm max}$ given by numerical optimization.

\subsection{Star-chain Model}
\label{sec:star_chain_model}
\begin{figure}
    \centering
    \includegraphics[width=0.5\textwidth]{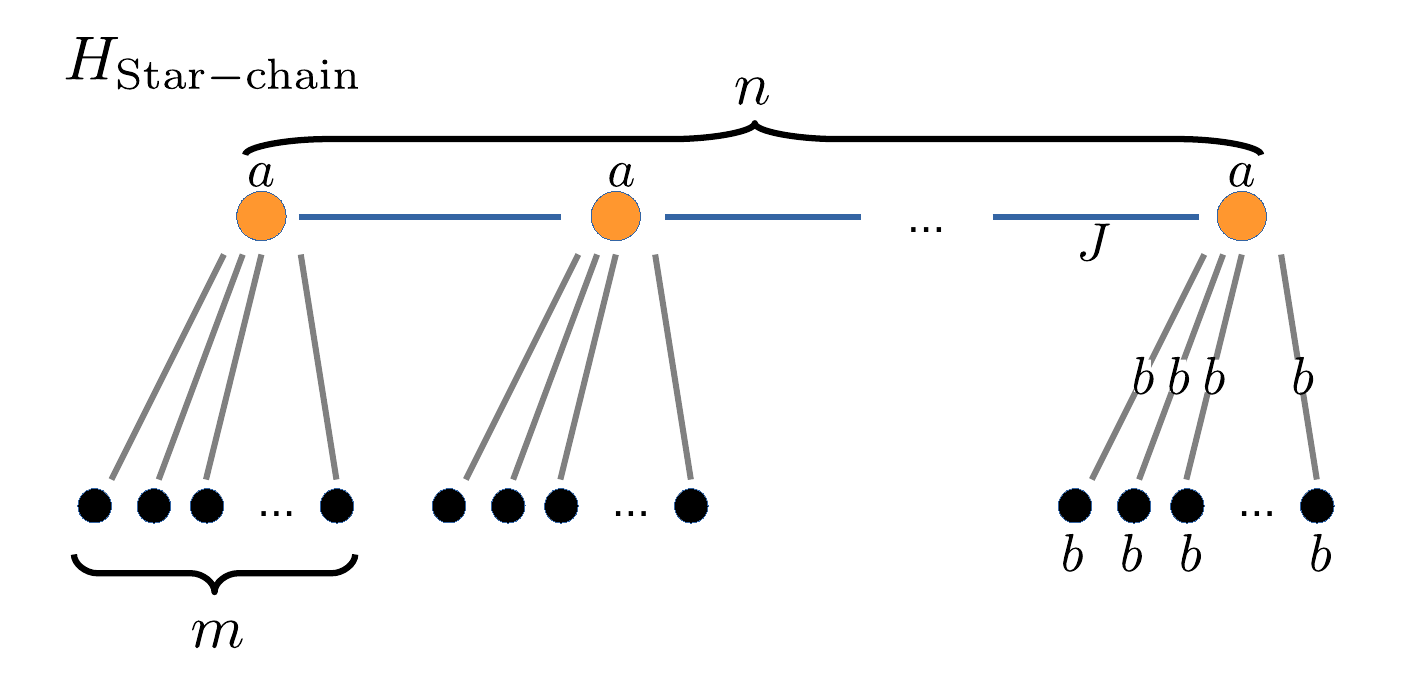}
    \caption{Representation of the Star-chain model~\eqref{eq:starchain}. The total number of spins is $N=n(m+1)$. The orange circles represent the $\alpha$-spins, coupled to each other through the blue lines, while the black circles represent the $(\alpha,i)$-spins coupled to their respective $\alpha$-spin through the gray lines. The values of the local fields $h_i$ and coupling terms $J_{ij}$ are reported in the Figure.}
    \label{fig:starchain}
\end{figure}

As the Star model arises from an unconstrained numerical optimization of $\C$ for Hamiltonians of the form~\eqref{eq:classical_H_generic} (cf. above Sec.~\ref{sec:star_model} and App.~\ref{sec:star_optimization}), we conjecture it to be the global optimum for such a class. However, the star-shaped connectivity of Eq.~\eqref{eq:H_star1} (Fig.~\ref{fig:star_scheme}) cannot be scaled to arbitrarily large number of constituents as it has long-range interactions. This motivates us to restrict the star-shaped connectivity to short-range interactions only, given rise to the hereafter named ``Star-chain model". Specifically, inspired by the Star model, we consider $N=n(m+1)$ spins as sketched in Fig.~\ref{fig:starchain}, described by the Hamiltonian
\begin{multline}
\label{eq:starchain}
 H_{\text{Star-chain}[N=m(n+1)]}(a,J,b):=\\
 a \sum_\alpha \sigma^z_\alpha + J \sum_\alpha \sigma^z_\alpha \sigma^z_{\alpha+1} + b \sum_{\alpha,i} (\sigma^z_\alpha+\mathbb{1}) \sigma^z_{\alpha,i}\;.
\end{multline}
Here $\alpha$ is the index identifying the central spin of each Star-like sub-unit (orange circles in the Figure), while $(\alpha,i)$ selects the $i$-th spin in each sub-unit (black circles), i.e.
\begin{align}
    \alpha =1,\dots,n\;, \quad 
    i =1,\dots,m \;.
\end{align}
Short-range interactions are guaranteed by considering a fixed value of $m$. The partition function of the Star-chain model  can be computed analytically (cf. App.~\ref{app:analytics_starchain}),
\begin{align}
    & \quad \quad \quad \quad \quad \quad Z_\text{Star-chain}=\lambda^{n}_{-}+\lambda^{n}_{+}\;,\\ \nonumber
    & \lambda_{\pm}\!=\! \frac{2^{m-1}}{AC} \!
    \left( C^2(A^2 \!+\! B^m)
    \!\pm\! \sqrt{4A^2 B^m \!+\! C^4(A^2 \!-\! B^m)^2}
    \right)\! ,
\end{align}
where $A=e^{\beta a}$, $B=\cosh{(2\beta b)}$, and $C=e^{\beta J}$. From the spectral point of view, this model guarantees a $2^{n_\downarrow m}$ degeneracy for each energy level with $n_{\downarrow}$ down $\alpha$-spins. In particular, when all the $\alpha$-spins are down, i.e. $\sigma^z_\alpha=-1\ \forall \alpha$, corresponding to a $2^{mn}$ degeneracy.
Moreover, if the couplings $J$ are negative and strong enough to force all the $n$ $\alpha$-spins to be the same (for a detailed analysis of the needed coupling strengths, see Sec.~\ref{sec:scaling-constraints} and App.~\ref{app:noise_tol}), one is left with two configurations only, $\sigma_\alpha= \pm 1 \ \forall \alpha$.
The case $\sigma_\alpha= 1$ corresponds to an evenly spaced, binomially distributed spectrum
$E=na + 2b \sum_{\alpha,i}\sigma_{\alpha,i}$,
while the case $\sigma_\alpha= -1$ corresponds to 
$E=-na$ with degeneracy $2^{nm}$.
%
It should be noticed that  these configurations
effectively lead to the same spectrum as the one of the Star model. More precisely, while the Star spectrum consists in $2^{N-1}$ states with binomial spectrum~\eqref{eq:spectrum_star_binom} and other $2^{N-1}$ states that are completely degenerate, see Eq.~\eqref{eq:spectrum_star_flat}, the Star-chain has, in the limit of large $-J$, $2^{mn}$ states with a binomial spectrum, $2^{mn}$ degenerate states, and $(2^{n}-2)2^{mn}$ remaining  arbitrarily high energy levels that can be neglected as justified earlier in this work. Property~2~\eqref{eq:prop2} then ensures that the Star-chain model can exhibit a heat capacity at least as large of that of a system having $1$ ground state and a $2^{mn}+mn$-fold degenerate first excited level.
This spectrum is achieved by choosing
\begin{align}
    -na=na-2bmn+4b \rightarrow b(mn-2)=na \;,
\end{align}
which leads to $\mathcal{C}^{\text{Star-chain}[N]}_{\rm max}\sim (\ln (2^{mn}+mn))^2/4$ 
This shows that $\C$ is essentially quadratic in $N=n(m+1)$, i.e. 
\begin{align}
\label{eq:abel_scaling}
    \mathcal{C}^{\text{Star-chain}[N]}_{\rm max}\gtrsim \left(\frac{m\ln 2 }{2(m+1)}\right)^2 N^2\;.
\end{align}
Equation~\eqref{eq:abel_scaling} makes clear how large values of $m$ increase the  achievable heat capacity, (see App.~\ref{app:opt_values_tables} for the optimal values of $\C$ and the corresponding Hamiltonian parameters). For $m=N-1$, the Star-chain model coincides with the Star model ($n=1$, cf. Figs.~\ref{fig:star_scheme} and~\ref{fig:starchain}). Let us note that short-range interactions impose a maximum $m$, but this one only impacts the prefactor of the quadratic scaling. Hence, it does not change the quadratic scaling of $\C$ demonstrated with the Star-chain model.

\subsection{Implementation in the Chimera graph}
\label{sec:chimera_embed}
\begin{figure}
    \centering
    \includegraphics[width=0.4\textwidth]{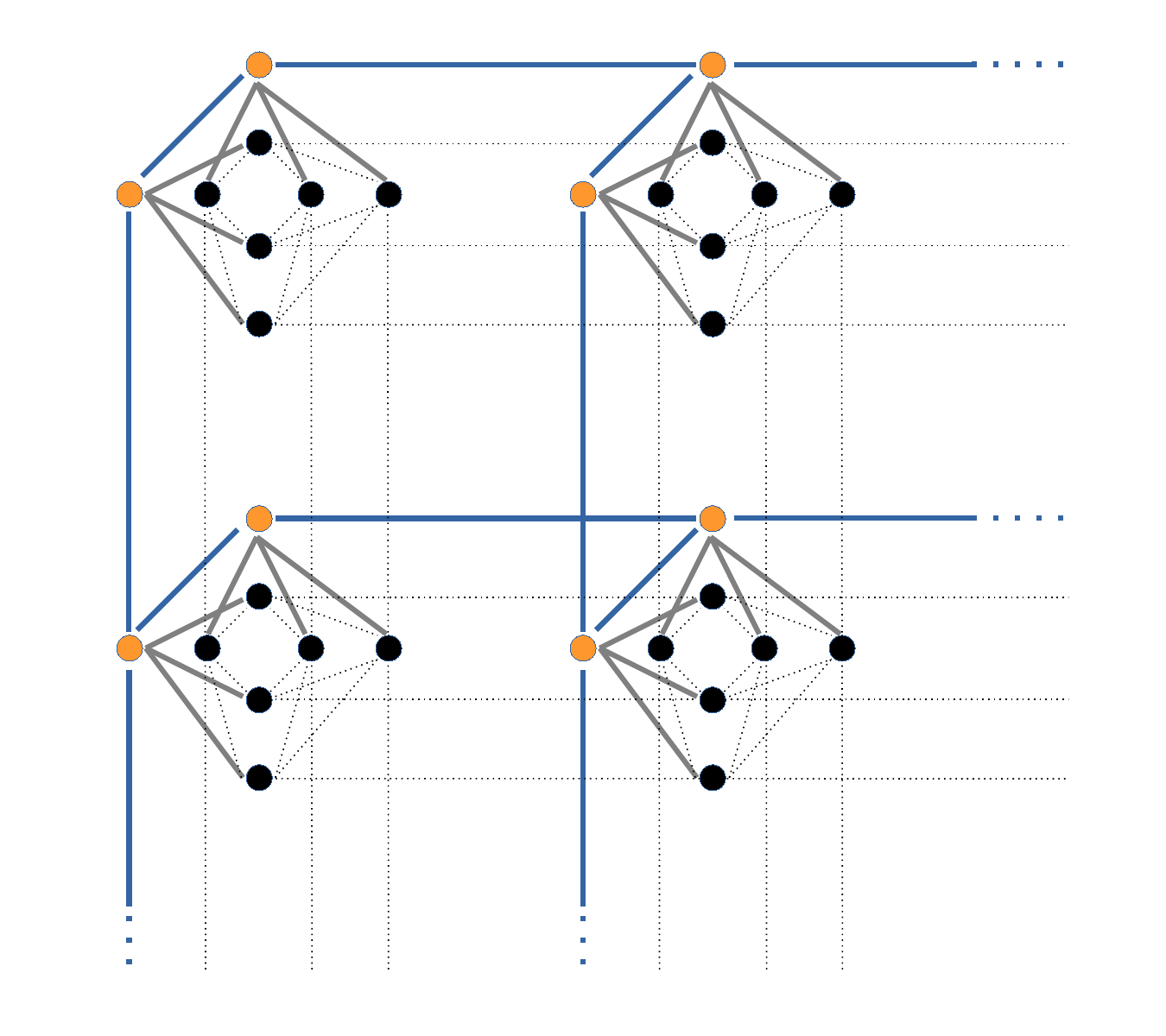}
    \caption{Embedding of the Star-chain model for $m=3$ (see Fig.~\ref{fig:starchain}) into the Chimera graph, which is used by D-Wave Systems~\cite{D-Wave}. As in Fig.~\ref{fig:starchain}, orange circles represent the $\alpha$-spins, coupled to each other through the blue lines, while black circles represent the $(\alpha,i)$-spins coupled to the respective $\alpha$-spin through the gray lines. Dotted lines represent unused couplings of the Chimera architecture (i.e. where $J_{ij}=0$).}
    \label{fig:embedding}
\end{figure}

Quantum annealers are devices governed by programmable quantum spin Hamiltonians,  therefore representing a natural platform to test our findings.
Interestingly, the topology of the interactions of the Star-chain model with  $m=3$ (cf. Fig.~\ref{fig:starchain}) can be embedded into the Chimera graph (cf. Fig.~\ref{fig:embedding}) of the D-Wave annealing quantum processor~\cite{D-Wave}.
This means that, as from~\eqref{eq:abel_scaling}, a programmable spin network in the Chimera graph can reach at least $\sim\frac{m^2}{(m+1)^2}=9/16$ of the ultimate bound $\mathcal{C}^{\rm opt}(2^N)$~\eqref{eq:optimal_scaling}. Remarkably, numerical optimization of $\C$ for the Chimera model results indeed in the Star-chain model with $m=3$ represented in Fig.~\ref{fig:embedding} (see App.~\ref{sec:app_dwave}). Notice also that there exists new architectures of the D-Wave annealers, such as the Pegasus graph~\cite{amin2015searching,dattani2019pegasus,boothby2020next-generation}, which can reach higher connectivities, and therefore higher values of $m$ for which the Star-chain Hamiltonian can be embedded. Such optimal thermometer probes could be used, for example, to precisely measure the surrounding effective temperature of the annealer (to be compared with the cryostat temperature), and overall to gain a better understanding of the D-Wave annealer as an open quantum system~\cite{benedetti2016estimation,marshall2019power,bian2020quantum,albash2021comparing}.

Let us emphasize a specificity of both the Star and Star-chain models, that may become relevant for practical applications. For both models, \emph{it is enough to measure a single spin} to perform temperature estimation. In the regime of large $N$, the only relevant energy levels contributing to the Gibbs state are the ground level, and the first excited level (higher excited levels are exponentially suppressed in the statistics, cf. App.~\ref{subsec:star_statistics} and previous discussions). We can distinguish between these two cases by simply measuring the value of $\sigma^z_1$ for the Star model~\eqref{eq:H_star1}, or \emph{any} of the $\sigma^z_\alpha$ in the Star-chain model~\eqref{eq:starchain}. 

\subsection{Scaling and constraints on the strength of the interactions}
\label{sec:scaling-constraints}
\begin{figure}[!tb]
	\centering
	\includegraphics[width=0.99\columnwidth]{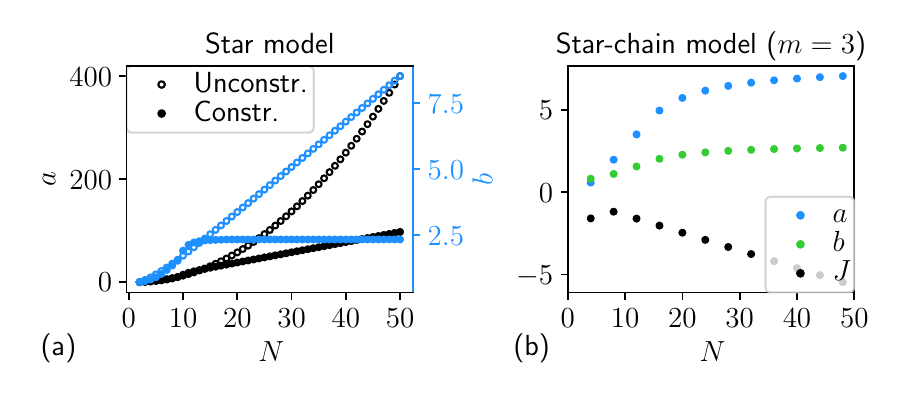}
\caption{Parameter scaling of the Star model (a) and Star-chain model for $m=3$ (b) in their configurations that maximise $\C$ (we obtain nearly identical values of $\C$ for both cases). In the ``unconstrained'' case (empty circles), $b$ increases linearly with $N$, and $a$ quadratically. This corresponds to the optimal choice $a=b(N-3)$ of Sec.~\ref{sec:star_model}. In the ``constrained'' case (full circles), it is possible to find solutions in which $b$ is limited by a constant, while $a$ increases linearly. These solutions preserve the same numerical value of $\mathcal{C}$, and are 
found optimizing the heat capacity over $a$ and $b$, and choosing as initial point for the optimization $b=2.2$ and $a = 2N-3$.}
	\label{fig:star_param_scaling}
\end{figure}

While the results presented above are very promising, one challenging requirement of the optimal configurations is the strength of the interactions between the constituents. This one becomes increasingly demanding for large $N$. For instance, engineering the optimal spectrum for the Star model with a first-excited state degeneracy $2^{N-1}+N-1$ requires the scaling $b\propto N$ and $a\propto N^2$ as $N$ grows (see ``unconstrained'' dots in Fig.~\ref{fig:star_param_scaling}(a)), accompanied by a relative precision $\propto N^{-2}$ for both parameters (cf. Sec.~\ref{sec:star_model} and App.~\ref{app:noise_tol}).

However, as we show in App.~\ref{app:analytics_star}, there exist solutions that are mathematically sub-optimal but numerically indistinguishable in terms of $\C$, with much more favorable scaling of the Hamiltonian parameters. In fact, even when limiting $b$ to be bounded by a constant, it is possible to achieve the desired quadratic scaling of $\C$, arbitrarily close to the optimal value Eq.~\eqref{eq:optimal_star_scaling}. These solutions feature a finite $b$, whose precise value becomes irrelevant, and a linear scaling of $a\propto N$, which admits a relative precision $\propto N^{-1}$ (see Apps.~\ref{app:noise_tol},~\ref{app:analytics_star} and ``constrained'' dots in Fig.~\ref{fig:star_param_scaling}(a)).
Similarly, the Star-chain model features solutions in which the scaling of its $a,b$ parameters is bounded, while $J$ scales linearly (see Fig.~\ref{fig:star_param_scaling}(b)). As seen in Fig.~\ref{fig:star_param_scaling}, for reasonable sizes of the thermal probe up to 50 spins, these scaling induce Hamiltonian parameters of moderate strength, both for the Star and Star-chain model.

\begin{figure}[!htb]
	\centering
	\includegraphics[width=0.99\columnwidth]{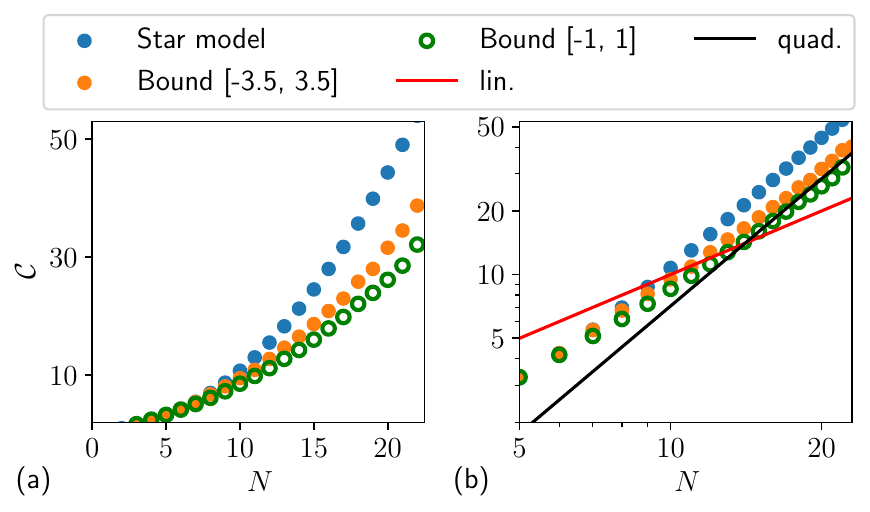}
\caption{Comparison of the heat capacity, as a function of $N$, on a linear scale (a), and on a log-log scale (b). The ``bound'' curves represent numerical maximizations of the general spin Hamiltonian~(\ref{eq:classical_H_generic}), where all parameters are constrained to be in a certain interval, i.e. $h_i, J_{ij}\in [-c,c]$, with $c$ shown in the legend. The red and black lines are reference quadratic and linear scalings.}
	\label{fig:star_3}
\end{figure}

Finally, it is possible to use the machine learning optimization method to maximize $\C$ over all possible Hamiltonians of the form Eq.~\eqref{eq:classical_H_generic} with the additional constraint for the parameters $\{h_i,J_{ij}\}$ to be bounded (See App.~\ref{app:bounded_params} for details). Our numerical optimization leads to configurations that are too complex and case-dependent to be discussed in generality. However, the resulting maximal heat capacities seem to indicate that a quadratic scaling is still possible under such constraints, see Fig.~\ref{fig:star_3}.

\section{Comparison to alternative models}
\label{sec:other_models}

\begin{figure*}[ht]
  \includegraphics[width=\textwidth]{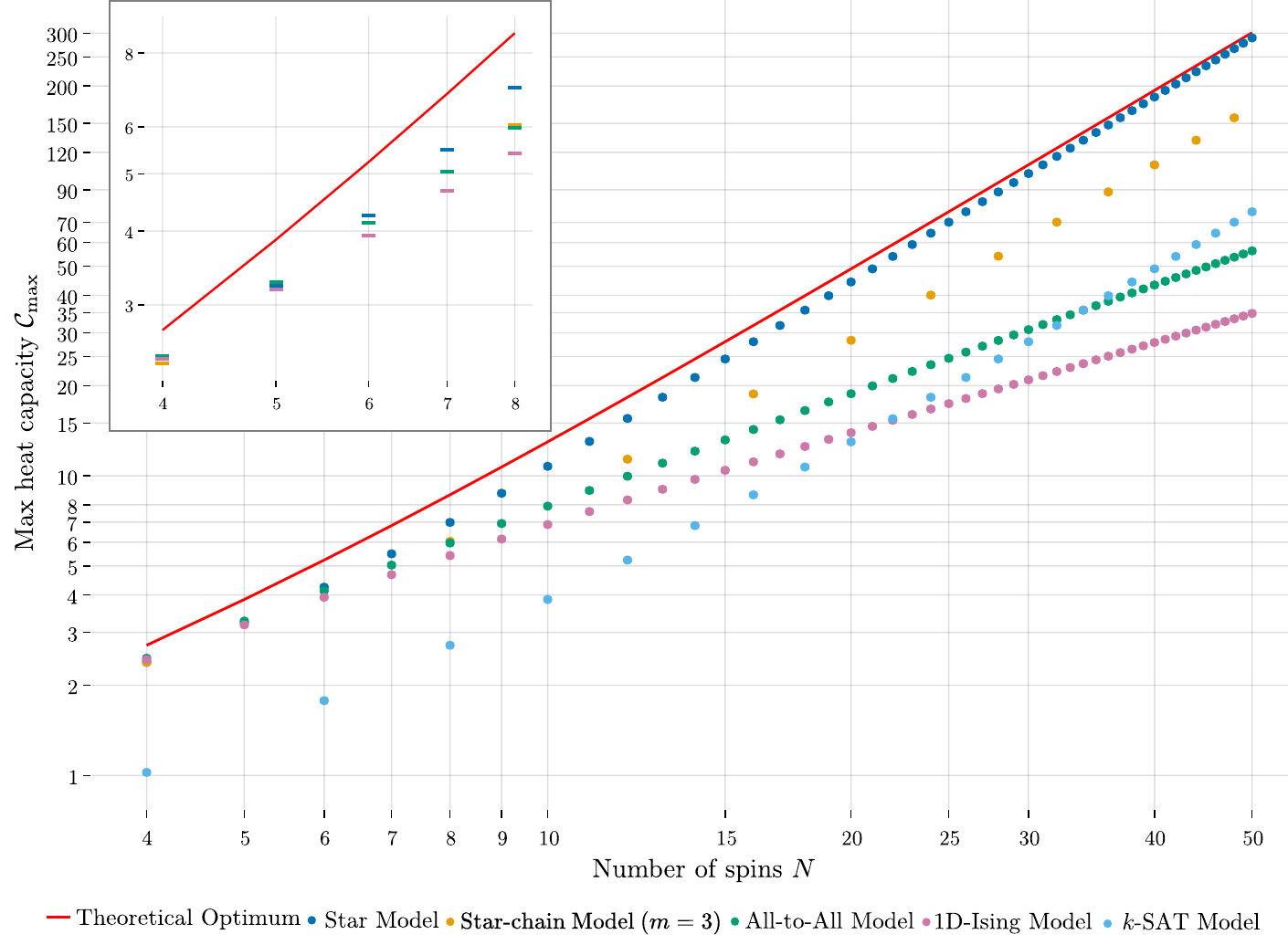}
  \caption{Detailed performance of the optimal spin-based thermometers found in this work. Our optimal architectures, ``Star model" and ``Star-chain model" demonstrate a quadratic $\propto N^2$ scaling of the maximal heat capacity $\C$ in terms of the number $N$ of total spins employed. This is to be compared with the extensive $\propto N$ scaling of standard models, such as the 1D Ising chain, or the All-to-all model described in the text. For $N\geq 6$, the Star model provides the highest heat capacity for Hamiltonians of the form~\eqref{eq:classical_H_generic} and can reach the mathematical bound $\mathcal{C}^{\rm opt}$~\eqref{eq:Copt_def} (red line in the plot) in the large $N$ limit. The Star-chain model has a similar behaviour while only using short-range interaction, and it can be programmed on current quantum annealers (cf. Sec.~\ref{sec:chimera_embed}). A detailed description and discussion of all the models we consider is given in the text.}
  \label{fig:bigscalingplot}
\end{figure*}

In Fig.~\ref{fig:bigscalingplot}, we compare the maximum values of the heat capacity $\C$ for different models of spin Hamiltonians, as the number of spins $N$ grows. The Star and Star-chain models show a quadratic scaling in $N$ that eventually surpasses all standard models - such as the Ising model in 1D, as well as a model of uniform ``all-to-all" interactions. The latter show instead the standard thermodynamic extensive scaling, i.e. linear in $N$, of the heat capacity. Below, we briefly describe each of the relevant alternative models to which our results obtained for the Star and Star-chain models have to be compared.

\paragraph{Ising Lattices.}
The 1D Ising model is arguably the simplest candidate for an interacting-spin thermometry probe. For $N$ spins, it is defined by the Hamiltonian 
\begin{equation}
    H_{1D} (\vec{h},\vec{J}) := -\sum_{i=1}^N h_i \sigma_i^z   - \sum_{i=1}^N  J_i \sigma_i^z  \sigma_{i+1}^z,
\end{equation}
where we choose periodic boundary conditions $\sigma_{N+1}\equiv\sigma_1$. The heat capacity for this model can be efficiently computed with standard techniques~\cite{mussardo2010statistical}. Numerically maximization leads consistently to homogeneous interactions $J_{i}=J$ and local fields $h_i=h$. As expected, an Ising chain probe will at most achieve a linear scaling in $N$ of the heat capacity, as seen in Fig.~\ref{fig:bigscalingplot}. Note that a 2-dimensional Ising model can achieve a slightly higher scaling at criticality, i.e. $\C_{\rm max}\propto N\ln N$~\cite{fisher1967theory,izmailian2002exact}, while the 3-dimensional Ising model has $\C_{\rm max}\propto N^{1.058}$ using critical scaling~\cite{campisi2016power}.

\paragraph{All-to-all symmetric model}
Another relevant model for this work is a model with all-to-all interactions, completely symmetric under permutations. Its Hamiltonian takes then the form
\begin{equation}
	H_\text{All}(h, J) := -h \sum_{i=1}^N \sigma_i^z   -J \sum_{i<j}  \sigma_i^z  \sigma_j^z\,.
	\label{eq:alltoall}
\end{equation}
It describes a complete graph with homogeneous interactions $J>0$ and local fields $h>0$. Taking the systems' symmetries into account, we get the following $\binom{N}{k}$-degenerate eigenenergies for $k\leq N$ up-spins:
\begin{equation}
	E_k = h(N-2k) +\frac{J}{2}\left[ 4k(N-k) - N(N-1)  \right].
\end{equation}
As shown in App.~\ref{app:alltoall_model}, the all-to-all model shows a ``large degeneracy" of the first excited level for small $N$. It is remarkable that for $N\leq 5$, the all-to-all model appears to have the highest heat capacity among the models we investigated  (see inset in Fig.~\ref{fig:bigscalingplot}) and consistently emerged from numerical optimisation in the same regime (cf. Sec.\ref{sec:star_optimization}). However, the degeneracy of the first excited state increases linearly in $N$, to be contrasted with the exponential increase of the Star and Star-chain models. This ultimately leads to a linear scaling of the heat capacity of the symmetric all-to-all model for large $N$.

\paragraph{k-SAT model and the exponential degeneracy}
Finally, we notice that in Refs.~\cite{Chancellor2016,Dodds2019}, a Hamiltonian replicating a global AND-operation between $M$ logical bits (represented by $M$ spins) was introduced, with the aid of $M$  ancillary spins. Such Hamiltonian was proposed as the basic element to build general models to solve $k$-SAT problems~\cite{impagliazzo2001on_the_complexity}. We will thus refer to it as the \emph{k-SAT model}. A logical AND identifies a single string (without loss of generality, the string given by $111\dots 1$, $M$ times) with an energy $E_{\rm AND}$ different from the energy $E_{\overline{\rm AND}}$ associated to all the other $2^M-1$ logical strings. 
Formally, this spectrum coincides with the ideal two-level degenerate model~\eqref{eq:deg_Ham}, and therefore the Hamiltonian proposed in \cite{Chancellor2016,Dodds2019} exhibits the desired 
quadratic scaling of the maximum $\C$, more precisely $\C_{\rm max}^{k-{\rm SAT}[N]}=\C^{\rm opt}(2^\frac{N}{2})$ (cf. Fig.~\ref{fig:bigscalingplot}). The construction uses a total $N=2M$ of spins (the neglected levels correspond to energies that can be made arbitrarily high, see~\cite{Chancellor2016}), that is, an overhead of $N/2$.
The Star and Star-chain models achieve similar degeneracies while using a much smaller overhead, i.e. a $1$-spin overhead for the Star model and a $N/(m+1)$-spins overhead for the Star-chain model. In Table~\ref{tab:overheads}, we compare these models in terms of the excited-level degeneracy and scaling of $\C$, as well as the locality of the interactions.

\begin{table}
\begin{tabular}{c|c|c|c}
Model & 1st excited deg. & Asymptotic $\C_{\rm max}$ & Short-range?\\
\hline \hline
$k$-Sat~\cite{Chancellor2016}   & $2^{\frac{N}{2}}-1$  & $\sim\dfrac{(\ln 2)^2}{4} \dfrac{N^2}{4}$ & $\xmark$  \\  \hline 
  Star   & $ 2^{N-1}+N-1$ & $\sim \dfrac{(\ln 2)^2}{4} (N-1)^2$  & $\xmark$ \\ \hline
  Star-chain & $ 2^{\frac{mN}{m+1}}+\dfrac{mN}{m+1}$ & $\sim\dfrac{(\ln 2)^2}{4} \dfrac{m^2 N^2}{(m+1)^2} $ & $\cmark$
\end{tabular}
\caption{Models recreating an effective spectrum with a single ground state and an exponentially degenerate first excited level. With the same total number $N$ of spins, the k-Sat model ``sacrifices'' half of them to obtain a $\sim 2^{\frac{N}{2}}$ degeneracy, while the Star model has a single spin overhead ($\sim 2^{N-1}$ degeneracy), and the Star-chain a $N/(m+1)$ overhead ($\sim 2^{\frac{mN}{m+1}}$ degeneracy). Moreover, the Star-chain model can be realised with short-range interactions.}
\label{tab:overheads}
\end{table}

\section{Conclusions and outlook}
\label{sec:conclusions}
In this work, we addressed the problem of maximizing the heat capacity $\C$ of physically realisable quantum systems, which amounts to engineer the best probe for temperature estimation in the context of equilibrium thermometry~\cite{mehboudi2019thermometry}.
Using a combination of  analytical derivations, Machine-Learning methods, and physical insights,
we explore the design space of spin Hamiltonians with two-body interactions and local magnetic fields, 
discovering Hamiltonians with star-shaped topology that can approach the theoretical maximum of $\C$ in the limit of large systems.
Additionally, we showed that an arbitrarily good approximation can be achieved  when requiring these interactions to be short-ranged.
The models emerging from such optimisation achieve a Heisenberg-like scaling of the sensitivity, without the use of entanglement, contrary to the well-known case of phase estimation in quantum optics~\cite{giovannetti2006quantum}. Remarkably, these models show a simple architecture of the interactions that make them ideal probes also for adaptive temperature estimation schemes~\cite{mehboudi2022fundamental,Jrgensen2022}. We further showed that the models we found can be embedded in currently available quantum annealers~\cite{D-Wave}, making them highly attractive both from a theoretical and experimental points of view. 
These results pave the way to the physical realization of ultra-sensitive spin-based thermometers, valid also for alternative experimental platforms such as cold atoms~\cite{Jepsen2020},  NV centers~\cite{Zhou2020}, and Rydberg atoms~\cite{Ebadi2021}. Of particular interest is the use of these engineered optimal spin-network thermal probes for ultracold gases~\cite{Bouton2020,Mehboudi2019,Mitchison2020,Planella2022,Khan2022}.  



In terms of Hamiltonian spectrum engineering, we showed that the essential requirement for an optimal thermal probe made of $N$ constituents is the presence of a single ground state and an exponential 
degeneracy of the first excited level. This effective two-level spectrum also appears in other problems, in which we speculate that our work might have application, such as protein folding modelling \cite{zwanzig1992levinthal,zwanzig1995simple}, adiabatic Grover's search~\cite{farhi1998analog,roland2002quantum,Allahverdyan2022},  energy based boolean computation~\cite{Chancellor2016}, and quantum heat engines~\cite{allahverdyan2013carnot,campisi2016power,abiuso2020optimal,abiuso2020geometric,cavina2021maximum}.

An interesting challenge for the future is to characterise the relaxation timescale $\tau_{\rm rel}$ of the optimal probes derived here, see also Refs.~\cite{allahverdyan2013carnot,Allahverdyan2022}. Due to critical slowdown, we expect a trade-off between large heat capacity and slowness of the relaxation process.  It hence remains a relevant open question if a similar Hamiltonian engineering can be performed taking as a figure of merit $\mathcal{C}/\tau_{\rm rel}$, which would also have important consequences  in the optimization of thermal engines~\cite{allahverdyan2013carnot,campisi2016power,abiuso2020optimal,abiuso2020geometric,cavina2021maximum}. At the same time, it is  worth emphasising that when time is a resource for thermometry~\footnote{In this case, full relaxation to equilibrium is clearly suboptimal, and optimal protocols take place when the probe is in an out-of-equilibrium state (see e.g. ~\cite{mehboudi2019thermometry}).},  optimal non-equilibrium protocols require the same effective two-level structure of the models presented in this work, as recently shown in~\cite{sekatski2021optimal}. Another challenge is to move beyond the weak coupling assumption behind \eqref{eq:GibbsState}, and consider the optimisation of thermometer probes for the more general mean force Gibbs state~\cite{Miller2018,Trushechkin2022,glatthard2023energy}.  



\acknowledgements
We thank Rosario Fazio for fruitful discussions.
PA is supported by “la Caixa” Foundation (ID 100010434, Grant No. LCF/BQ/DI19/11730023), and by the Government of Spain (FIS2020-TRANQI and Severo Ochoa CEX2019-000910-S), Fundacio Cellex, Fundacio Mir-Puig, Generalitat de Catalunya (CERCA, AGAUR SGR 1381).
FN gratefully acknowledges funding by the BMBF (Berlin Institute for the Foundations of Learning and Data -- BIFOLD), the European Research Commission (ERC CoG 772230) and the Berlin Mathematics Center MATH+ (AA1-6, AA2-8). PAE gratefully acknowledges funding by the Berlin Mathematics Center MATH+ (AA1-6). 
GH and MPL acknowledge funding from the Swiss National Science Foundation through a starting grant PRIMA PR00P2\_179748 and an Ambizione Grant No. PZ00P2-186067, and through the NCCR SwissMAP.

\section*{Code availability}
The code  used to generate these results can be provided upon request to the authors.

\bibliography{BIB.bib}

\newpage
\widetext

\appendix 

\section{ADAM optimization and the emergence of the Star (and Star-chain) models}
\label{sec:star_optimization}
In this Appendix we explain how we carried out the numerical optimization of the heat capacity using methods that are commonly employed in machine learning.

Let us consider an arbitrary Hamiltonian $H(\theta)$ that depends on a set of parameters $\theta$. The $\theta$ parameters could be, for example, the $h_i$ and $J_{ij}$ parameters in Eq.~(\ref{eq:classical_H_generic}). Our aim is to determine the value of the parameters $\theta$ that maximize the heat capacity of the system. As discussed in the main text, this is equivalent to maximizing the Hamiltonian variance of the thermal state $\Delta^2_\beta H$ given in Eq.~(\ref{eq:varianceDiag}).

In machine learning, it is common to minimize a ``loss function'' $\mathcal{L}(\theta)$ that depends on a set of parameters. One way to determine the value of $\theta$ that minimizes $\mathcal{L}(\theta)$ is to use gradient descent. This consists of starting from a random value of the parameters $\theta$, computing the gradient $\partial_\theta \mathcal{L}(\theta)$, and updating the parameters according to
\begin{equation}
    \theta_\text{new} = \theta_\text{old} -\alpha \partial_\theta \mathcal{L}(\theta),
    \label{eq:gd}
\end{equation}
where $\alpha>0$ is the so-called ``learning rate'' that determines how large of a step we take in parameter space in the opposite direction of the gradient. If $\alpha$ is small enough and $\mathcal{L}(\theta)$ is differentiable, then it is guaranteed that $\mathcal{L}(\theta_\text{new})\leq \mathcal{L}(\theta_\text{old})$; reiterating this gradient descent step many times, we will reach a local minimum.

However, this method is prone to getting stuck in local minima, may take many iterations to converge, and choosing appropriate values of $\alpha$ is not always straightforward. An alternative to the update rule in Eq.~(\ref{eq:gd}) is given by ADAM (Adaptive Moment Estimation) \cite{kingma2014}; this method was empirically found to converge better in a variety of problem. As Eq.~(\ref{eq:gd}), it only requires the calculation of the gradient at each iteration, but it improves over it in various ways, so we refer to Ref.~\cite{kingma2014} for details.

In order to find the parameters $\theta$ that maximize the heat capacity of the system described by $H(\theta)$, we use as loss function
\begin{equation}
    \mathcal{L}(\theta) \equiv - \Delta^2_\beta H,
\end{equation}
such that minimizing the loss function corresponds to maximizing the Hamiltonian variance. We then start from a random choice of $\theta$ and use the ADAM optimization method to minimize the loss function. We compute the gradient of the variance using backpropagation \cite{goodfellow2016}, which is a common machine learning algorithm that automatically computes the gradient of a function in a given point. In particular, we use the PyTorch framework to compute the Hamiltonian variance  of the thermal state, its gradient, and to perform the ADAM optimization using the default hyperparameters. 

We now display some of the results we found with this method in different classes systems.

\subsection{$N$ Spin Hamiltonian}
\label{appsub:adam_at_various_N}

In this subsection we show how the Star model emerged from the numerical optimization considering the $N$ spin Hamiltonian given in Eq.~(\ref{eq:classical_H_generic}), and we provide some details on the optimization method. The optimization was carried out as described above considering $\{h_i\}$ and $\{J_{ij}\}$ as the $\theta$ parameters. Both are initialized randomly between $-1$ and $0$. We performed separate optimizations for $N\in \{2,3,\dots, 15\}$, finding the All-to-All model for $N\in\{ 2,\dots,6 \}$, and the Star model for $N \geq 7$. For $N\in \{2,\dots,11\}$, the results were found running a single optimization with fixed learning rate $\alpha= 0.001$ for $60000$ steps (although most optimizations converge much sooner). However, for $N\geq 12$, this choice would sometimes get stuck in local minima. For $N=\{12,13\}$, we ran the optimization multiple times as detailed above, and chose the model with the largest heat capacity. For $N=\{14,15\}$, to avoid getting stuck in local minima, we used a common technique in Machine Learning, which consists of scheduling the learning rate, i.e. of varying it at each optimization step. In particular, we used the ``CyclicLR'' scheduler of PyTorch that varies the learning rate in a triangular fashion between a minimum $\alpha_\text{min}$ and a maximum $\alpha_\text{max}$ value. For $N=14$, we chose $\alpha_\text{min}=0.0003$ and $\alpha_\text{max}=0.2$, and halved the amplitude of the triangle at every repetition (such that, asymptotically, the learning rate converges to $\alpha_\text{min})$. The number of steps during the ``up phase'' of the triangle was chosen to be $6000$. For $N=15$, we chose $\alpha_\text{min}=0.001$ and $\alpha_\text{max}=0.1$ without halving the amplitude of the triangle at every repetition. Also in this case the ``up phase'' consists of $6000$ steps.

\begin{figure}[!htb]
	\centering
	\includegraphics[width=0.63\columnwidth]{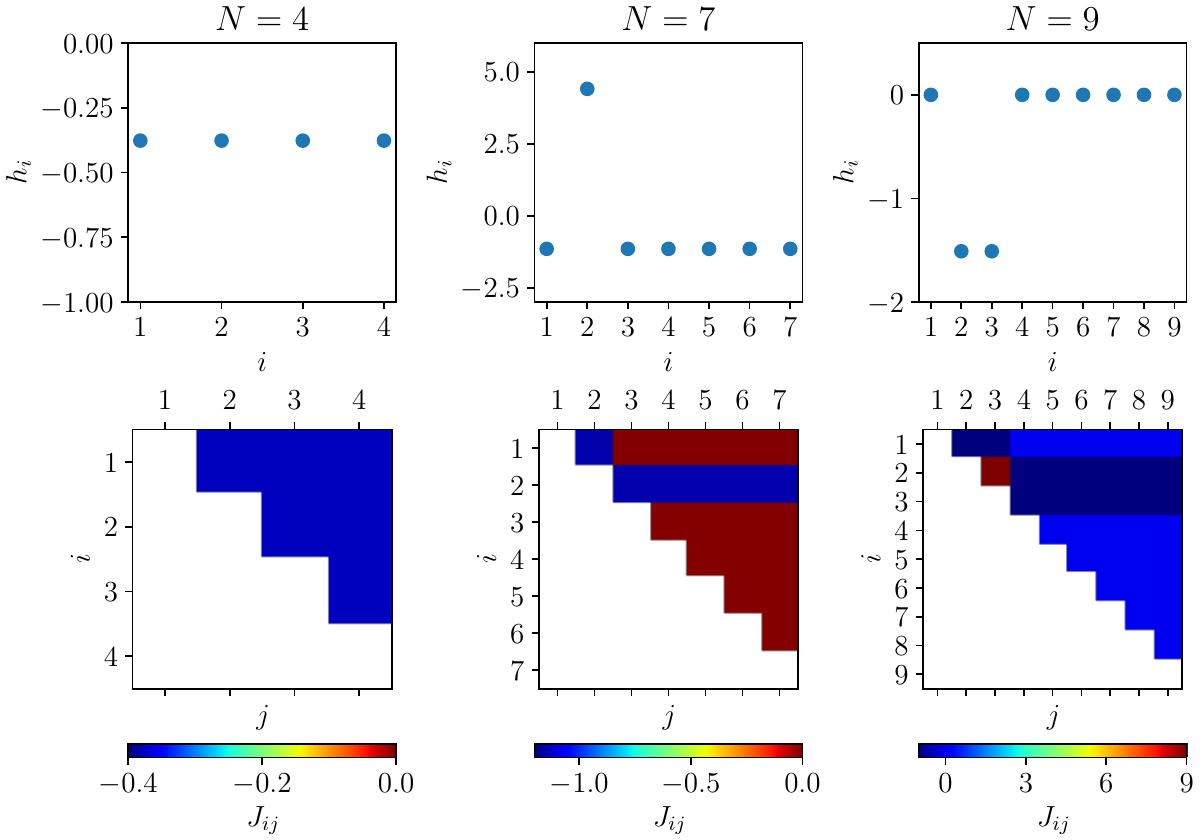}
\caption{Values of $h_i$ and of $J_{ij}$ found with our numerical method for $N=4$ (left panels), $N=7$ (middle panels), and $N=9$ (right panels). The upper panels show $h_i$ as a function of the site index $i=1,\dots,N$, while the lower panels show the value of $J_{ij}$ (the color) as a function of the site indices $i$ and $j$. Since $J_{ij}$ is only defined for $i<j$, a white square is shown when such condition is not satisfied.}
	\label{fig:classical_spins_params}
\end{figure}
To show the emergence of the Star model, in Fig.~\ref{fig:classical_spins_params} we show the values of $h_i$ and of $J_{ij}$ found with our numerical method for $N=4$ (left panels), $N=7$ (middle panels), and $N=9$ (right panels). The upper panels show $h_i$ as a function of the site index $i=1,\dots,N$, while the lower panels show the value of $J_{ij}$ (the color) as a function of the site indices $i$ and $j$. Since $J_{ij}$ is only defined for $i<j$, a white square is shown when such condition is not satisfied.

\begin{figure}[!htb]
	\centering
	\includegraphics[width=0.2\columnwidth]{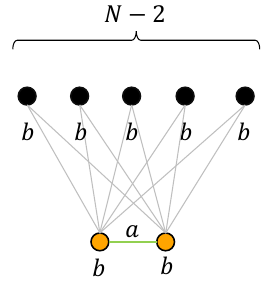}
\caption{Schematic representation of the Hamiltonian $H_{\overline{\text{Star}}(N)}(a,b)$ given in Eq.~(\ref{eq:star_2}).}
	\label{fig:star_2}
\end{figure}

As we can see, for $N=4$ all parameters take the same value ($h_i=J_{ij}\approx -0.377$ for every $i$ and $j$); this corresponds to the All-to-All model described in Eq.~(\ref{eq:alltoall}) with $h=J\approx 0.377$. For $N=7$, we see that all spins are interchangeable except for a single privileged spin corresponding to $i=2$. Indeed, $h_2\approx 4.41$, while $h_{i\neq 2} \approx -1.15$, and $J_{ij}$ is non null, and equal to $-1.15$, only when $i$ or $j$ are equal to $2$. This is precisely the Star model as described in Eq.~(\ref{eq:H_star1}) with $a \approx 4.41$ and $b \approx -1.15$. For $N=9$, we find a model where all spins are interchangeable, except for two privileged spins corresponding to $i=2,3$. Indeed, $h_{i}=0$ except for $h_2=h_3\approx -1.51$. Also $J_{ij}=0$ when both $i$ and $j$ are not $2$ or $3$, while $J_{23}\approx 8.96$, and $J_{ij}=-1.51$ when $i$ or $j$ are $2$ or $3$, but not both. The Hamiltonian of this model can be written as
\begin{equation}
    H_{\overline{\text{Star}}(N)}(a,b) =  a \,\sigma_1^z  \sigma_2^z +b \sum_{i=1,2} \sigma_i^z  \left(\mathbb{1} + \sum_{j=3}^N \sigma_j^z\right),
    \label{eq:star_2}
\end{equation}
and schematically represented as in Fig.~\ref{fig:star_2} with $a\approx 8.96$ and $b\approx=-1.51$.
Interestingly, it can be seen that $H_{\overline{\text{Star}}(N)}(a,b)$ has the same exact spectrum as $H_{{\text{Star}}(N)}(a,b)$, and therefore the same heat capacity. Therefore, while they are physically two different models, they have identical characteristics as thermometers. 

\begin{figure}[!htb]
	\centering
	\includegraphics[width=0.47\columnwidth]{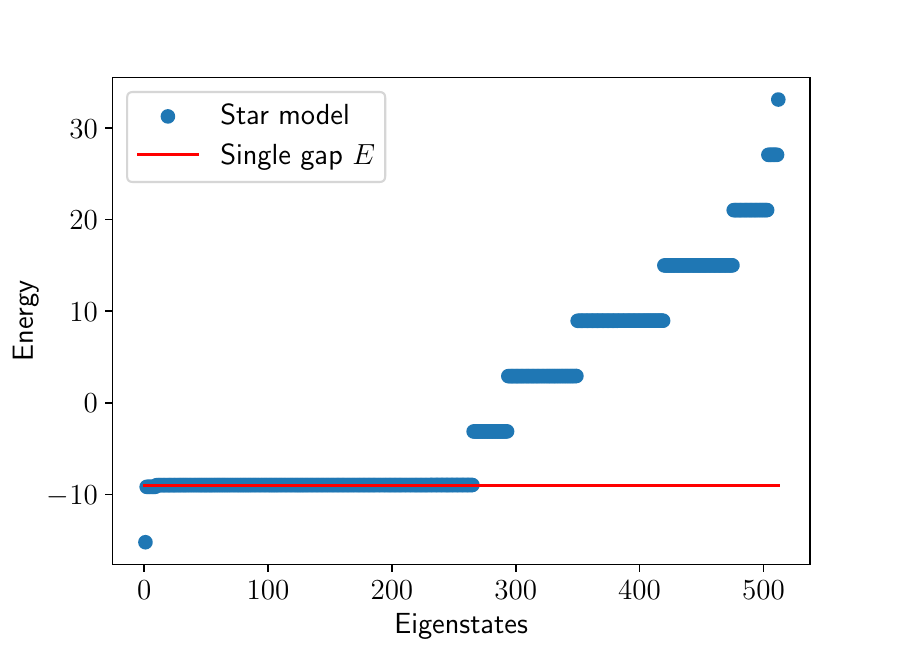}
\caption{Spectrum of the Star model, found with the numerical optimization for $N=9$, represented as dots with the corresponding energy on the y-axis. The red line represents the energy $E$ (measured from the ground state energy) that maximizes the heat capacity of the degenerate Hamiltonian $H_{\rm deg}$ [see Eq.~(\ref{eq:deg_Ham})] for $N=9$.}
	\label{fig:star_spectrum}
\end{figure}
At last, in Fig.~\ref{fig:star_spectrum} we analyze the spectrum of the Star model corresponding to $N=9$. The individual eigenenergies are plotted as dots on the y-axis. For comparison, we plot as a red line the value of $E$ (measured from the ground state energy) that maximizes the heat capacity of the degenerate Hamiltonian $H_{\rm deg}$ [see Eq.~(\ref{eq:deg_Ham})] for $N=9$.
As expected, there is a single ground state, a highly degenerate first excited state (with a degeneracy approximately given by half of the states), and further excited states with a binomial degeneracy. Interestingly, and as suggested by the Lemma of App.~\ref{app:Lemma_proof}, the value of $E$ is quite similar to the energy of the highly degenerate first excited state.

\subsection{$N$ Spin Hamiltonian with bounded parameters}
\label{app:bounded_params}
In this subsection we discuss how we performed the optimization of the heat capacity of $N$ spins with an additional bound on the magnitude of the parameters that lead to the ``bound'' curves in Fig.~\ref{fig:star_3}. In particular, we wish to maximize Eq.~(\ref{eq:classical_H_generic}) with respect to $h_i$ and $J_{ij}$, with the additional constraint that
\begin{align}
    |h_i| &\leq c, & |J_{ij}|\leq c,
    \label{eq:h_const}
\end{align}
where $c>0$ is a real constant.

Since our method works well for unconstrained optimizations, we introduce the following parameterization
\begin{align}
    h_i &= c\tanh x_i, & J_{ij}&= c\tanh y_{ij},
    \label{eq:h_tan}
\end{align}
where $x_i$ and $y_{ij}$ are real parameters. Since the hyperbolic tangent produces values in $[-1,1]$, the parameterization of Eq.~(\ref{eq:h_tan}) guarantess to satisfy the constraint in Eq.~(\ref{eq:h_const}) for any value of $x_i$ and $y_{ij}$.

We therefore apply the same optimization described above, but choosing $\{x_i\}$ and $\{y_{ij}\}$ as our unconstrained optimization parameters $\theta$, instead of $\{h_i\}$ and $\{J_{ij}\}$.

In particular, the orange and green dots in Fig.~\ref{fig:star_3} were produced the following way. We initialize the $\{x_i\}$ and $\{y_{ij}\}$ parameters randomly in the interval $[-1.5, 1.5]$. Then, for each value of $N \in \{3,4, \dots,20\}$, we repeat the optimization $12$ times, choosing the one with the highest heat capacity. In particular, $6$ repetitions are performed with learning rate $\alpha = 0.01$, and $6$ with $\alpha=0.03$. For $N\in\{21,22\}$, we do the same but choosing as learning rates respectively $\alpha=0.01$ and $\alpha=0.003$.

\subsection{Optimal values for the Star model and the Star-chain.}
\label{app:opt_values_tables}
In this subsection we provide some details regarding the heat capacity maximization in the Star and Star-chain models. In particular, in Tables~\ref{tab:params_1} and \ref{tab:params_2} we provide the explicit values plotted in Fig.~\ref{fig:star_param_scaling}. 

All three optimizations are carried out using the Adam optimizer for $6000$ steps and backpropagation to compute the gradients as described in App.~\ref{sec:star_optimization}, but we only optimize over the $a$ and $b$ parameters of the Star model, and over the $a$, $b$ and $J$ parameters of the Star-chain model.

In particular, the ``unconstrained'' case of the Star model was optimized fixing the condition $a=b(N-3)$, and optimizing only over $b$. The initial value is set to $b=6$, and the learning rate is set at $\alpha=0.01$. In the ``constrained'' case of the Star model, we optimize over $a$ and $b$ choosing as initial values $a=2N-3$ and $b=2.2$, and learning rate $\alpha=0.001$. In the Abel model with $m=3$, we optimize over $a$, $b$ and $J$. For $N=4$, we choose as initial values $a=3.5$, $b=1.55$ and $J=-1.6$. For higher values of $N$, we choose as initial values the parameters that maximize the previous optimization. We set the learning rate at $\alpha=0.003$.

\begin{table}[h]
\begin{tblr}{
  colspec = {|c||c|c||c|c||c|c|c|},
  hspan = even, 
  cell{1}{2} = {c=2}{c}, 
  cell{1}{4} = {c=2}{c}, 
  cell{1}{6} = {c=3}{c}, 
}
\hline
 & Star model (unconstr.) & & Star model (constr.) & & Star-chain model & & \\ \hline
$N$ & $a$& $b$& $a$& $b$& $a$& $b$& $J$ \\
\hline\hline
2 & -0.711 & 0.711 & -0.711 & 0.711 & - & - & - \\
3 & 0.000 & 0.797 & -0.136 & 0.752 & - & - & - \\
4 & 0.894 & 0.894 & 0.578 & 0.810 & 0.578 & 0.810 & -1.600 \\
5 & 2.015 & 1.007 & 1.518 & 0.897 & - & - & - \\
6 & 3.398 & 1.133 & 2.797 & 1.020 & - & - & - \\
7 & 5.070 & 1.267 & 4.482 & 1.173 & - & - & - \\
8 & 7.052 & 1.410 & 6.553 & 1.340 & 1.964 & 1.101 & -1.191 \\
9 & 9.358 & 1.560 & 9.041 & 1.520 & - & - & - \\
10 & 11.998 & 1.714 & 13.722 & 1.905 & - & - & - \\
11 & 14.977 & 1.872 & 17.489 & 2.123 & - & - & - \\
12 & 18.297 & 2.033 & 20.311 & 2.216 & 3.504 & 1.559 & -1.612 \\
13 & 21.960 & 2.196 & 22.760 & 2.263 & - & - & - \\
14 & 25.967 & 2.361 & 25.032 & 2.289 & - & - & - \\
15 & 30.318 & 2.527 & 27.208 & 2.304 & - & - & - \\
16 & 35.013 & 2.693 & 29.322 & 2.314 & 4.953 & 2.021 & -2.038 \\
17 & 40.053 & 2.861 & 31.397 & 2.320 & - & - & - \\
18 & 45.438 & 3.029 & 33.444 & 2.324 & - & - & - \\
19 & 51.168 & 3.198 & 35.473 & 2.326 & - & - & - \\
20 & 57.243 & 3.367 & 37.490 & 2.328 & 5.720 & 2.267 & -2.468 \\
21 & 63.664 & 3.537 & 39.495 & 2.329 & - & - & - \\
22 & 70.431 & 3.707 & 41.493 & 2.329 & - & - & - \\
23 & 77.543 & 3.877 & 43.488 & 2.329 & - & - & - \\
24 & 85.001 & 4.048 & 45.477 & 2.329 & 6.164 & 2.411 & -2.903 \\
\hline
\end{tblr}
\caption{Values of the parameters of the Star and Star-chain models, that optimize the heat capacity, plotted in Fig.~\ref{fig:star_param_scaling} (see caption of Fig.~\ref{fig:star_param_scaling}) for details. The second half of the parameters are given in Table~\ref{tab:params_2}.}
\label{tab:params_1}
\end{table}

\begin{table}[h]
\begin{tblr}{
  colspec = {|c||c|c||c|c||c|c|c|},
  hspan = even, 
  cell{1}{2} = {c=2}{c}, 
  cell{1}{4} = {c=2}{c}, 
  cell{1}{6} = {c=3}{c}, 
}
\hline
 & Star model (unconstr.) & & Star model (constr.) & & Star-chain model & & \\ \hline
$N$ & $a$& $b$& $a$& $b$& $a$& $b$& $J$ \\
\hline\hline
25 & 92.805 & 4.218 & 47.465 & 2.329 & - & - & - \\
26 & 100.956 & 4.389 & 49.451 & 2.329 & - & - & - \\
27 & 109.452 & 4.560 & 51.436 & 2.329 & - & - & - \\
28 & 118.294 & 4.732 & 53.420 & 2.329 & 6.452 & 2.504 & -3.336 \\
29 & 127.483 & 4.903 & 55.404 & 2.329 & - & - & - \\
30 & 137.018 & 5.075 & 57.388 & 2.330 & - & - & - \\
31 & 146.899 & 5.246 & 59.371 & 2.329 & - & - & - \\
32 & 157.127 & 5.418 & 61.354 & 2.328 & 6.649 & 2.568 & -3.767 \\
33 & 167.701 & 5.590 & 63.339 & 2.329 & - & - & - \\
34 & 178.621 & 5.762 & 65.324 & 2.329 & - & - & - \\
35 & 189.887 & 5.934 & 67.310 & 2.329 & - & - & - \\
36 & 201.500 & 6.106 & 69.295 & 2.329 & 6.790 & 2.614 & -4.195 \\
37 & 213.460 & 6.278 & 71.280 & 2.329 & - & - & - \\
38 & 225.766 & 6.450 & 73.267 & 2.329 & - & - & - \\
39 & 238.418 & 6.623 & 75.254 & 2.329 & - & - & - \\
40 & 251.417 & 6.795 & 77.241 & 2.329 & 6.896 & 2.648 & -4.622 \\
41 & 264.762 & 6.967 & 79.230 & 2.329 & - & - & - \\
42 & 278.453 & 7.140 & 81.218 & 2.329 & - & - & - \\
43 & 292.492 & 7.312 & 83.206 & 2.329 & - & - & - \\
44 & 306.876 & 7.485 & 85.196 & 2.329 & 6.979 & 2.676 & -5.047 \\
45 & 321.607 & 7.657 & 87.187 & 2.330 & - & - & - \\
46 & 336.685 & 7.830 & 89.176 & 2.330 & - & - & - \\
47 & 352.109 & 8.002 & 91.167 & 2.330 & - & - & - \\
48 & 367.879 & 8.175 & 93.158 & 2.330 & 7.046 & 2.697 & -5.472 \\
49 & 383.996 & 8.348 & 95.150 & 2.330 & - & - & - \\
50 & 400.460 & 8.520 & 97.141 & 2.330 & - & - & - \\
\hline
\end{tblr}
\caption{Continuation of the parameters displayed in Table~\ref{tab:params_1}.}
\label{tab:params_2}
\end{table}

\subsection{Quantum $N$ Spin Hamiltonian}
\label{app:quantum}
In this subsection, we employ our numerical optimization method to maximize the heat capacity of the most generic two-body spin Hamiltonian, namely
\begin{equation}
    \bar{H}_\text{quantum} = \sum_{i \atop \mu\in\{x,y,z\}} \bar{h}^{(\mu)}_i \sigma^{\mu}_i + \sum_{i < j \atop \mu,\nu\in\{x,y,z\}} \bar{J}^{(\mu,\nu)}_{ij} \sigma^{\mu}_i \sigma^{\nu}_j,
\end{equation}
where $\bar{h}^{(\mu)}_i$ and $\bar{J}^{(\mu,\nu)}_{ij}$ are arbitrary parameters. Since the heat capacity only depends on the spectrum of the Hamiltonian, we can perform arbitrary unitary operations to $\bar{H}_\text{quantum}$ without changing its spectrum, thus its heat capacity. Choosing local unitary transformations of the form
\begin{equation}
    U_i = \exp\left[ i \sum_{\mu\in \{x,y,z\} } \theta_\mu \sigma^\mu_i  \right],
\end{equation}
where $\theta_\mu$ are three suitable angles,
we can always rotate $\sum_{\mu\in \{x,y,z\}} \bar{h}^{(\mu)}_i \sigma^{\mu}_i$ into an operator proportional only to $\sigma^z_i$. Therefore, applying the appropriate unitary transformation on each spin site, we obtain the Hamiltonian
\begin{equation}
    {H}_\text{quantum} = \sum_{i} {h}_i \sigma^{z}_i + \sum_{i < j \atop \mu,\nu\in\{x,y,z\}} {J}^{(\mu,\nu)}_{ij} \sigma^{\mu}_i \sigma^{\nu}_j,
    \label{eq:quantum}
\end{equation}
where ${h}_i$ and ${J}^{(\mu,\nu)}_{ij}$ are arbitrary parameters.

In this subsection, without loss of generality, we optimize Eq.~(\ref{eq:quantum}) considering ${h}_i$ and ${J}^{(\mu,\nu)}_{ij}$ as optimization parameters $\theta$. We performed a separate optimization for $N\in \{2,3,\dots,9\}$ with a fixed learning rate $\alpha=0.001$, performing $20000$ optimization steps, and starting from random initial values of the parameters uniformly distributed between $0$ and $1$. In all cases, we found values of the heat capacity that are identical (up to numerical errors) to the values found considering the $N$ spin Hamiltonian with only $\sigma^z$, i.e. the model, given by Eq.~(\ref{eq:classical_H_generic}), considered in the previous subsection. Furthermore, these solutions also have the same spectrum found in the previous subsection. However, they are not the same model: indeed, the optimal values of ${J}^{(\mu,\nu)}_{ij}$ that we find are non-zero when $\mu\neq z$ and $\nu\neq z$. This can be understood in the following way: since the spectrum and the heat capacity are invariant under unitary transformations, we can apply any unitary transformation to the Star model to generate different models that display the same spectrum and heat capacity. Therefore, there is an infinitely large class of systems with the same optimal heat capacity, and our optimization method converges randomly to one of these solutions.

\begin{figure}[!htb]
	\centering
	\includegraphics[width=0.8\columnwidth]{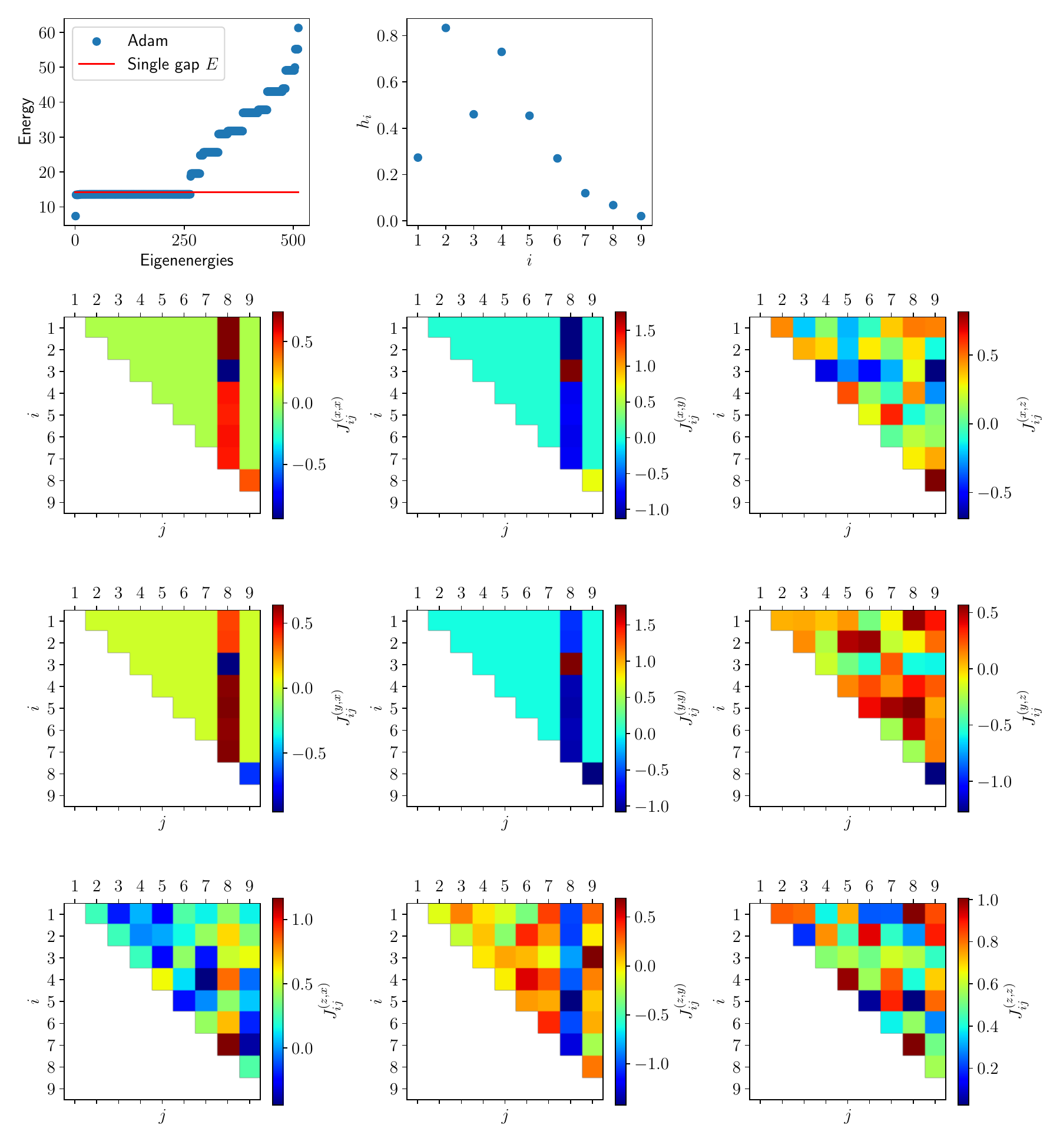}
\caption{Optimization results for the quantum spin Hamiltonian in Eq.~(\ref{eq:quantum}) for $N=9$. The first row shows the spectrum and the $h_i$ as in Figs.~\ref{fig:star_spectrum} and \ref{fig:classical_spins_params}. Each of the lower 9 panels displays $J^{(\mu,\nu)}_{ij}$, for all combinations of $\mu,\nu\in\{x,y,z\}$, as a function of the site index $i$ and $j$ (similar to Fig.~\ref{fig:classical_spins_params}). Since $J^{(\mu,\nu)}_{ij}$ is only defined for $i<j$, a white box is plotted when such condition is not fulfilled.}
	\label{fig:quantum_spins_params}
\end{figure}
As an example, in Fig.~\ref{fig:quantum_spins_params} we plot the spectrum, $h_i$, and ${J}^{(\mu,\nu)}_{ij}$ (for $\mu,\nu\in \{x,y,z\}$), that we found for $N=9$, in the same style as in Figs.~\ref{fig:classical_spins_params} and \ref{fig:star_spectrum}. As we can see, there is some structure in $J^{(\mu,\nu)}_{ij}$ for $\mu,\nu\in\{x,y\}$ that privileges a specific spin index (number $8$ in this case). However, it is clear that this model is different from the $N$ spin Hamiltonian with only $\sigma^z$. Nonetheless, we see that the spectrum, and thus the heat capacity, is essentially the Star spectrum (compare the first panel of Fig.~\ref{fig:quantum_spins_params} with Fig.~\ref{fig:star_spectrum}). The very small discrepancies are due to the numerical optimization method that reached parameters near the local minima, but not exactly. As previously anticipated, the ``noisyness'' that is visible in many panels can be explained by the infinite number of models that yield the same spectrum, such that the numerical method converges to a random one based on the initial stochastic choice of the parameters.

\subsection{D-Wave annealer Hamiltonian}
\label{sec:app_dwave}
In this subsection we consider a spin Hamiltonian as in Eq.~(\ref{eq:classical_H_generic}) with only $\sigma^z_i$ terms, but we constrain the optimization to reflect the topology of the interactions of D-Wave annealers. In particular, we consider the Chimera graph as in Fig.~\ref{fig:embedding}, and we focus on 3 units, i.e. $24$ spins. This corresponds to excluding the lower right unit of Fig.~\ref{fig:embedding}. Mathematically, we enforce elements of $J_{ij}$ to be null whenever a connection between spin $i$ and $j$ is not present in the topology, and then we minimize the loss function considering the non-null $\{J_{ij}\}$ and $\{h_i\}$ parameters as $\theta$. We use the Star optimization method for $20000$ steps at a fixed learning rate, and randomly initializing the parameters between $-1.5$ and $1.5$. In fact, we ran the optimization $3$ times: once with $\alpha=0.01$, and twice with $\alpha=0.03$. This yielded values of the heat capacity between $39.99$ and $41.57$.

\begin{figure}[!htb]
	\centering
	\includegraphics[width=0.8\columnwidth]{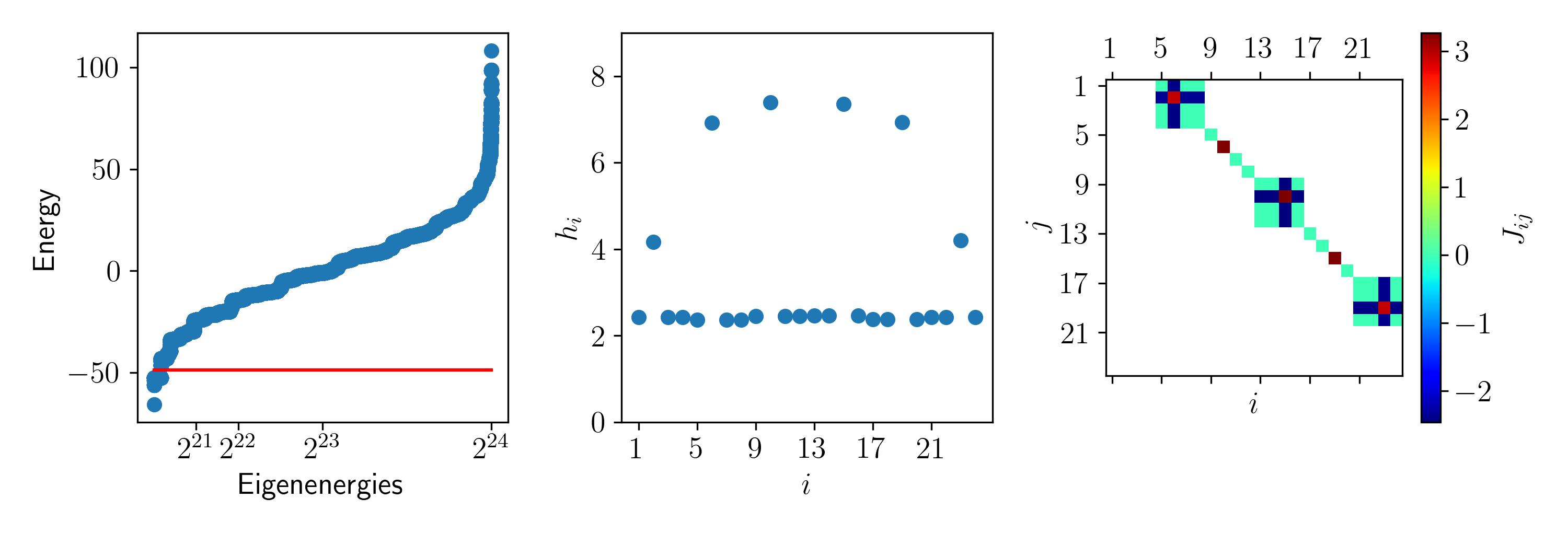}
\caption{Optimization results for the Chimera graph topology of the interactions of D-Wave annealers with $3$ units ($24$ spins). 
The first panel shows the spectrum of the model as in Fig.~\ref{fig:star_spectrum}, while the second and third panels show the values of $h_i$ and of $J_{ij}$ as in Fig. \ref{fig:classical_spins_params}. Since $J_{ij}$ is only defined for $i<j$ and when spins $i$ and $j$ are coupled according to the Chimera graph shown in Fig.~\ref{fig:embedding}, a white box is plotted whenever $J_{ij}$ is not defined.}
	\label{fig:chimera_params}
\end{figure}

In Fig.~\ref{fig:chimera_params} we show the numerical results that we found in the optimization run that yielded the largest heat capacity (corresponding to $\alpha=0.03)$. The first panel shows the spectrum in the same style as Fig.~\ref{fig:star_spectrum}, while the second and third panels show the values of $h_i$ and $J_{ij}$ as in Fig.~\ref{fig:quantum_spins_params}. To better understand the results, we applied a local unitary flip of $\sigma_i^z$ in all sites where $h_i<0$ (which does not change the spectrum, thus the heat capacity). This amounts to changing the sign of $h_i$ whenever $h_i$ is negative, and correspondingly changing the sign of $J_{ij}$ and $J_{ji}$ for all $j$. The indexing of the spins is such that the first unit is described by $i\in \{1,\dots,8\}$, the second by $i\in \{9,\dots,16\}$, and the third by $i\in \{17,\dots,24\}$. Furthermore, spins $\{5,\dots,8\}$ of the first unit are coupled to spins $\{9,\dots,12\}$ of the second unit, and spins $\{13,\dots,16\}$ of the second unit are coupled to spins $\{17,\dots,20\}$ are the third unit (see non-white boxes in the last panel of Fig.~\ref{fig:chimera_params}).

As we can see, this model is very similar to the Star-chain model with $m=3$ embedded into the Chimera graph as in Fig.~\ref{fig:embedding}. Indeed, there are two privileged spins per unit (corresponding to spins $2,6$ in the first unit, $10,15$ in the second, and $19,23$ in the third). These are represented in orange in Fig.~\ref{fig:embedding}. These spins have a larger on-site potential as compared to all other ones (see middle panel of Fig.~\ref{fig:chimera_params}), and they are each coupled to $3$ spins within the same unit (see the ``dark blue crosses'' in the last panel of Fig.~\ref{fig:chimera_params}). Furthermore, the three units are linked to each other through these privileged spins (see the two brown isolated dots in the last panel of Fig.~\ref{fig:chimera_params}).

\section{A small Lemma of (Property 2)}
\label{app:Lemma_proof}
In this Appendix we prove a theoretical Lemma that leads to Property 2 in the main text, Sec.~\ref{sec:spectra_properties}. 
The Lemma considers two Hamiltonians, $H_1$ and $H_2$, such that $H_1$ has a single ground state and a $k_1$-degenerate excited state ($k_1+1$ levels in total), while $H_2$ has the same spectrum and additional $k_2$ excited states above (totaling $1+k_1+k_2$ levels) that is,
\begin{align}
    H_1&=0\ket{0}\bra{0}+\sum_{i=1}^{k_1}  E \ket{i}\bra{i}\;,\\
    H_2&=0\ket{0}\bra{0}+\sum_{i=1}^{k_1}  E \ket{i}\bra{i}+ \sum_{\alpha=k_1+1}^{k_1+k_2} E_\alpha \ket{\alpha}\bra{\alpha}\;,
\end{align}
with $0\leq E\leq E_\alpha\; \forall\alpha$.
Consider now the realistic situation in which these Hamiltonians are controlled via internal coupling parameters $\vec{\lambda}$, such as is the case of our work.
The Lemma has two assumptions: \textit{i)} it is possible to control the first excited gap $ E(\vec{\lambda})$, contemporary to keeping the additional $\alpha$-levels above \textit{ii)} $E_\alpha(\vec{\lambda})\geq E(\vec{\lambda})$. A simple scenario in which these assumptions are satisfied is the simple Hamiltonian $ H_2(\lambda)=\lambda\left(\sum_{i=1}^{k_1}  E \ket{i}\bra{i}+ \sum_{\alpha=k_1+1}^{k_1+k_2} E_\alpha \ket{\alpha}\bra{\alpha}\right)$.
Under these assumptions, the Lemma states that the maximal achievable heat capacity with $H_2$ is always larger than the maximum heat capacity obtainable with $H_1$.
\begin{align}
    \max_{E}\C(H_1)\leq \max_{E\leq E_\alpha} \C(H_2)\;.
\end{align}

\paragraph*{Proof of the Lemma.}
When computing the variance of the energy in a thermal state, global shifts in the energy do not matter. For this reason we rewrite the same Hamiltonians putting the $k_1$  $\ket{i}$ levels to zero, i.e.
\begin{align}
    H'_1&=- E\ket{0}\bra{0}\;,\\
    H'_2&=- E\ket{0}\bra{0}+ \sum_{\alpha} E'_\alpha \ket{\alpha}\bra{\alpha}\;.
\end{align}
with $ E\geq 0$ and $E_\alpha-E\equiv E'_\alpha\geq 0$. \\
We now use temperature units $\beta=1$, to simplify the discussion. The thermal states are therefore
\begin{align}
    \rho^{(1)} =\frac{e^{-H'_1}}{\Tr[e^{-H'_1}]}\;,
    \quad
    \rho^{(2)} =\frac{e^{-H'_2}}{\Tr[e^{-H'_2}]}\;.
\end{align}
Let us call $p_0^{(1)}$ and $p_0^{(2)}$ the corresponding ground state populations,
\begin{align}
\label{eq:p_012}
 p_0^{(1)}=\dfrac{1}{1+k_1 e^{-E}}\;, \quad   
 p_0^{(2)}=\dfrac{1}{1+k_1 e^{-E}+\sum_\alpha e^{-(E'_\alpha+E)}}\;.
\end{align}
Notice that they both depend on the value of $E$, which we omit in the following for simplicity.
It is easy to compute the heat capacity (equivalently, the energy variance) as
\begin{align}
  \Delta^2 H_1= E^2\, p_0^{(1)}(1-p_0^{(1)})\;.
\end{align}
For what concerns $H_2$ instead (calling $p_\alpha$ the population of the level $E_\alpha$)
\begin{align}
\label{eq:OB}
    \Delta^2 H_2=
     E^2 p_0^{(2)} + \sum_\alpha  E'^2_\alpha p_\alpha - \left( - E\, p_0^{(2)} + \sum_\alpha  E'_\alpha p_\alpha \right)^2
     =E^2 p_0^{(2)}(1-p_0^{(2)}) + A + B\;,
\end{align}
with
\begin{align}
    A:=
    \sum_\alpha  E'^2_\alpha p_\alpha - \left(\sum_\alpha  E'_\alpha p_\alpha \right)^2 \geq 0\;,
   \quad
    B:= 
     2 E p_0^{(2)} \left(\sum_\alpha  E'_\alpha p_\alpha\right) \geq 0 \;.
\end{align}
where the inequalities follow from $B$ being trivially positive, while $A$ corresponds to the variance of an Hamiltonian having levels $ E_\alpha$ with population $p_\alpha$ and all the rest of the population $1-\sum_\alpha p_\alpha$ being at an energy $=0$.
It follows that
\begin{align}
\label{eq:h2_geq}
   \Delta^2 H_2 \geq 
     E^2 p_0^{(2)}(1-p_0^{(2)})\;.
\end{align}

Finally, let's compare the maximal value of the heat capacity in the two cases. Let's call $\bar{ E}^{(1)}$ the optimal value for the Hamiltonian $H_1$, which induces a ground state population equal to $\bar{p}^{(1)}_0=p^{(1)}_0(\bar{ E}^{(1)})$, i.e.
\begin{align}
    \max \Delta^2 H_1=\max_ E  E^2 p_0^{(1)}( E)(1-p_0^{(1)}( E))
    =(\bar{E}^{(1)})^2\bar{p}^{(1)}_0 (1-\bar{p}^{(1)}_0)\;.
\end{align}
It suffices to conclude now by noticing that $p_0^{(2)}(E)$ is an increasing function of $ E$ and is always smaller than $p_0^{(1)}(E)$, cf. Eq.~\eqref{eq:p_012}. This means that 
\begin{align}
    p_0^{(2)}(x)=p_0^{(1)}(y)\quad
    \text{implies}\quad
    x>y\;.
\end{align}
Therefore one can choose $p_0^{(2)}( E)=\bar{p}^{(1)}_0$ which will lead to $ E>\bar{ E}$ and therefore from~\eqref{eq:h2_geq}
\begin{align}
    \max \Delta^2 H_2 \geq(\bar{E}^{(1)})^2\bar{p}^{(1)}_0 (1-\bar{p}^{(1)}_0)\;,
\end{align}
concluding the proof.

\section{Parametric scaling and noise-tolerance}
\label{app:noise_tol}
In this section we estimate the strength and the precision that is needed in the engineering of the Hamiltonian parameters~\eqref{eq:classical_H_generic} in order to achieve the optimal Heisenberg-like scalings~(\ref{eq:optimal_scaling},\ref{eq:optimal_star_scaling},\ref{eq:abel_scaling}) of the models described in the main text. For simplicity, we will work in adimensional units in which $\beta=1$.

\subsection{Bandwidth tolerance in the degenerate model}
\label{subsec:noise_degmod}
As we argued in Sec.~\ref{sec:optimal_spectra}, the main property that optimal models satisfy in order to reach the optimal $\propto N^2$ scaling of the maximal heat capacity, is that of generating a single ground state and an exponentially-degenerate first excited level, i.e. approximating the degenerate Hamiltonian~\eqref{eq:deg_Ham} in the best possible way, as well as possibly having additional higher energy levels (cf. Lemma in App.~\ref{app:Lemma_proof}). However, in any physical realization, the resulting spectrum will have imperfections, when compared to~\eqref{eq:deg_Ham}. In particular, the exponentially-degenerate level might split into a bandwidth, or the overall gap might be shifted. Here we estimate the noise tolerance to such imprecisions in the first excited level.

\paragraph{Uniform shift.}
Consider first the case in which there is no splitting in the $D-1$ first excited levels, but a uniform error, that is
\begin{align}
    H=0\ket{0}\bra{0}+ E \sum_{i=1}^{D-1} \ket{i}\bra{i}\;,
\end{align}
and the value of $E$ is not exactly the optimal energy gap. We show here that as far as the imprecision does not scale with the dimension, the heat capacity behaves smoothly. We know, infact, that the optimal value of $E$, for large dimensions is $E\sim \ln (D-1)$. (cf. main text and~\cite{correa2015individual}). Suppose now that
\begin{align}
    E=(1+\epsilon)\ln(D-1)\equiv (1+\epsilon)\ln d
\end{align} 
Then, the resulting adimensional variance of the flat Hamiltonian with $d\equiv D-1$ degenerate excited states and gap $E$ can be computed given the ground state probability
\begin{align}
    p_0=\frac{1}{1+d e^{-E}}\;, \quad p_i=\frac{1}{d}(1-p_0) \; \text{for } i=1,\dots, d\;. 
\end{align}
It follows that
\begin{align}
\nonumber
\langle H^2\rangle - \langle H\rangle^2 =
\frac{de^{-E}E^2}{1+de^{-E}}-\frac{d^2e^{-2E}E^2}{(1+de^{-E})^2}=&
(\ln d)^2(1+\epsilon)^2\Big(\frac{d^{-\epsilon}}{1+d^{-\epsilon}}-\frac{d^{-2\epsilon}}{(1+d^{-\epsilon})^2}\Big)\\
=&(\ln d)^2\frac{(1+\epsilon)^2}{4\cosh^2(\frac{\epsilon}{2}\ln d)}\;.
\label{eq:robustness_degmod}
\end{align}
The leading term of the asymptotic energy variance, for large dimension $D-1=d$ is given, for $\epsilon\rightarrow 0$, by $(\ln d)^2/4$, i.e. recovering the results of~\cite{correa2015individual} (see also main text, Eq.~\eqref{Eq:UltimateLimit} and Sec.~\ref{sec:optimal_spectra}). Moreover, from expression~\eqref{eq:robustness_degmod} we see immediately that in order to keep such scaling, the denominator needs to be suppressed such that $\epsilon\ln d$ is bounded. This implies that the relative noise tolerance of the energy gap is given by
\begin{align}
    \epsilon \propto \frac{1}{\ln d} \sim \frac{1}{N}\;,
\end{align}
where we used that, in the case of $N$ constituents the dimension of the system is exponential in $N$. That is, the relative error $\epsilon$ should scale as $1/N$. Given $\ln d\propto N$, this is equivalent to the absolute error being bounded by a constant
\begin{align}
    E-\ln d \sim \mathcal{O}(1)\;.
\end{align}

\paragraph{Single eigenstate shift and bandwith tolerance.}
We now analyse the case in which the highly degenerate first excited level splits into separate energies. That is, consider a $D$-dimensional Hamiltonian with a single ground state and $d\equiv D-1$ levels contained in a bandwith $\delta$, i.e. (w.l.o.g. we can shift the ground-state energy to be negative, and the degenerate bandwith to be centered around 0)
\begin{align}
    H=-E \ket{0}\bra{0}+\sum_{i=1}^d E_i \ket{i}\bra{i} \quad \text{with} \quad -\delta \leq E_i \leq \delta \; \forall i \text{ and } E \simeq \ln d\;.
\end{align}
Computing the energy variance, we obtain
\begin{multline}
    \langle H^2\rangle - \langle H \rangle = E^2 p_0+\sum_i E_i^2 p_i -\left( E p_0+\sum_i E_i  p_i\right)^2 \\
    = E^2 p_0(1-p_0) + \sum_i E_i^2 p_i - \left(\sum_i E_i p_i\right)^2 - 2 E p_0 \sum_i E_i p_i
    = E^2 p_0(1-p_0)+ A+B\;,
\end{multline}
with
\begin{align}
  A=& \sum_i E_i^2 p_i - \left(\sum_i E_i p_i\right)^2\;,\\
  B=&  - 2 E p_0 \sum_i E_i p_i\;.
\end{align}
Notice now that the term $A$ is always positive, while the first term, $E^2 p_0(1-p_0)$ is the leading term in the degenerate model in which (cf. main text and~\cite{correa2015individual})
\begin{align}
    E\sim\ln d\propto N\;,\quad p_0\sim \frac{1}{2}\;.\quad \Delta^2 H\sim \frac{(\ln d)^2}{4}\propto N^2\;.
\end{align}
It is then easy to show that, similarly to~\eqref{eq:robustness_degmod}, as far as the $E_i$ levels are small and do not scale with the dimension, the optimal $N^2$ scaling is preserved.
This is easily seen as $A\geq 0$, while $B$ in case $-\delta \leq E_i\leq \delta$ is bounded as
\begin{align}
    |B|\leq 2 E\delta \sim \mathcal{O}(N)\;.
\end{align}
It follows that the leading term, $E^2p_0(1-p_0)\sim \mathcal{O}(N^2)$ remains dominant and the heat capacity achieves the $N^2$ scaling, as far as $p_0(1-p_0)$ is finite.
This is guaranteed by the fact that
\begin{align}
    p_0=\frac{1}{1+\sum_i e^{-(E+E_i)}}=\dfrac{1}{1+d^{-1}\sum_i e^{-E_i}}\;,
\end{align}
and therefore
\begin{align}
    \frac{1}{1+e^{\delta}}\leq p_0\leq  \frac{1}{1+e^{-\delta}}\;.  
\end{align}

\subsection{Consequences for the Star and Star-chain}
In the above subsection we estimated the noise tolerance of the energy spectrum of the degenerate Hamiltonian~\eqref{eq:deg_Ham} in order for the heat capacity to be close to its optimal value and maintain the Heisenberg-like scaling $\propto N^2$. The results indicate that the error in the spectral engineering should be constant while $N$ (and therefore the dimension $D$) grows. In terms of relative precision, as the optimal spectrum has a first excited gap $E\propto N$, this means a $1/N$ relative precision in the engineering of the spectrum around the optimal values.

However, the spectrum of the Hamiltonian is a function of its parameters $\{h_i,J_{ij}\}$~\eqref{eq:classical_H_generic}. In this subsection, we analyse what precision is needed in our main models, $H_{\rm Star}$~\eqref{eq:H_star1} and $H_\text{Star-chain}$~\eqref{eq:starchain} and how the estimation of the noise tolerance is reflected in the Hamiltonian parameters.

\paragraph{Reduction to $H_{\rm Star}$.}
\label{subsecapp:reduction}

First, for generic considerations, we notice that we limit ourselves to estimate the noise tolerance in the Star model only, because there is an exact mapping between the Star-chain and the Star model, in the limit of strong couplings $-J$, i.e. given (we allow a small relaxation in the Star-chain model, i.e. we assume different magnetic fields $a_\alpha$ on the $\sigma^z_\alpha$ spins, which is useful in the following analytical derivation, but does not significantly change the spectral properties of the model)
\begin{align}
   H_{\text{Star-chain}[N=n(m+1)]}(a_\alpha,J,b):=
 \sum_\alpha a_\alpha \sigma^z_\alpha + J \sum_\alpha \sigma^z_\alpha \sigma^z_{\alpha+1} + b \sum_{\alpha,i} (\sigma^z_\alpha+\mathbb{1}) \sigma^z_{\alpha,i}\;,
\end{align}
in the limit of high $J$, as explained in the main text, only configurations in which $\sigma^z_\alpha=\sigma^z_{\alpha'}$ are allowed, and therefore the effective Hamiltonian spectrum, up to an irrelevant global shift, becomes
\begin{align}
\label{eqapp:starchain_Jlim}
     H_{\text{Star-chain}[N=n(m+1)]}(a_{\alpha},J\rightarrow -\infty,b)=
  h \tilde{\sigma}^z  + b \sum_{\alpha,i} (\tilde{\sigma}^z+\mathbb{1}) \sigma^z_{\alpha,i}\;,
\end{align}
where $\tilde{\sigma}^z$ is a formal spin that has value $+1$ when $\sigma^z_\alpha=+1 \;\forall\alpha$ (similarly for $-1$), and $h=\sum_\alpha a_\alpha$\;.
The above~\eqref{eqapp:starchain_Jlim} formally coincides with the Star model~\eqref{eq:H_star1}
\begin{align}
  H_{\text{Star}[N=nm+1]}(a,b) &:= a \,\sigma_1^z + b \sum_{i=2}^{nm+1}  \left( \sigma_1^z + \mathbb{1}\right) \sigma_i^z    \;,
\end{align}
if one identifies $h\rightarrow a$\;.
Moreover the mapping preserves the parametrization in $b$, while $h:=\sum a_\alpha$ in the gets mapped to $a$ in the above equation. One could therefore assume $a_\alpha=0$ to be zero for all $\alpha$s except one, $h=a_{\bar{\alpha}}$, such that the mapping between the two models is complete and the parameterization is formally the same by mapping $a_{\bar{\alpha}}\rightarrow a$\;.

\paragraph{Parametric  scaling an noise-tolerance of $H_{\rm Star}$ in the optimal-degeneracy configuration.}
\label{par:noise_optimal_conf}
We thus consider here the $H_{\rm Star}$ Hamiltonian~\eqref{eq:H_star1}
\begin{align}
\label{eqapp:H_star1}
  H_{\text{Star}[N]}(a,b) &:= a \,\sigma_1^z + b \sum_{i=2}^{N}  \left( \sigma_1^z + \mathbb{1}\right) \sigma_i^z    \;,
\end{align}
As mentioned in the main text, by choosing $b\geq 0$ and $b(N-3)\leq a \leq b(N-1)$, it is ensured the presence of a single ground state at energy $E_0$, a $2^{N-1}$-degenerate level at energy $E_{\rm deg}$ and a 2nd excited, $N-1$-degenerate level $E_1$ as
\begin{align}
    E_0 \leq & E_{\rm deg} \leq E_1\;,\\
    E_0 =& a-2b(N-1)\;,\\
    E_{\rm deg}=& -a\;,\\
    E_1 =& a-2b(N-3)\;.
\end{align}
The optimal degeneracy of $2^{N-1}+N-1$ is reached when $a=b(N-3)$, but it is not necessary for the model to achieve its $N^2$ scaling of for the heat capacity $\C$. For simplicity, consider the choice $a=b(N-3)$. The first excited gap is, in this case
\begin{align}
    E_1-E_0=E_{\rm deg}-E_0=4b\;.
\end{align}
This means that, for such choice of parameters, in the asymptotic limit of large $N$, one has
$4b\sim  N \ln 2$\;, and consequently $a= b(N-3)\sim N(N-3)\frac{\ln 2}{4}$. That is, $b$ has a linear scaling in $N$ and $a$ has a quadratic scaling in $N$. This happens even if we relax the assumption of $a=b(N-3)$. In that case, it remains valid that
\begin{align}
    4b=E_1-E_0 \geq E_{\rm deg}-E_0 \sim E \ln 2\;,
\end{align}
therefore $b$ scales \emph{at least} linearly in $N$ and $a$ at least quadratically, as it satisfies $b(N-3)\leq a \leq b(N-1)$.

For what concerns the parametric error-tolerance for $a$ and $b$ in $H_{\rm Star}$, notice that we estimated above the gap-error tolerance, which results to be constant (cf.~\ref{subsec:noise_degmod}),
that is, one should have
\begin{align}
    2b(N-1)-2a=E_{\rm deg}-E_0\sim N\ln 2 + \delta
\end{align}
with $|\delta|=\mathcal{O}(1)$ bounded by a constant. From the above expression is easy to see that, if treated as independent, $b$ can have an error $\delta_b\sim \mathcal{O}(\frac{1}{N})$, while for $a$ the admitted error is $\delta_a\sim \mathcal{O}(1)$. It follows that both the relative error for $a$ and $b$ is inversely quadratic
\begin{align}
\label{eqapp:ab_tol}
    \frac{\delta_a}{a}\sim \mathcal{O}(N^{-2})\;, \quad \frac{\delta_b}{b}\sim \mathcal{O}(N^{-2})\;.
\end{align}

\paragraph{Noise-tolerance in the degeneracy-suboptimal configurations.}
\label{par:noise_suboptimal_conf}
In the subsection above, we considered the case in which the Star model~\eqref{eqapp:H_star1} is forced in its optimal configurations satisfying an exponentially-degenerate first excited level, i.e. $E_0\leq E_{\rm deg}\leq E_1$. We saw that this imposes a linear scaling on $b$ and quadratic scaling on $a$, and a relative error of order $\mathcal{O}(N^{-2})$ on both parameters. However, it is possible to show, as we do in Appendix~\ref{app:analytics_star}, that (slightly) suboptimal solutions exist, featuring a bounded value of $b$ $\forall N$, while still achieving quadratic scaling of $\C$. In fact, these solutions can have $\C$ to be  arbitrarily close to its optimal value $\C^{\rm Star}_{\rm max}$~\eqref{eq:optimal_star_scaling}).
While referring the reader to  Appendix~\ref{app:analytics_star} for the details, it is enough for our purposes to notice that, in such solutions, $b$ can take any finite value larger than a certain treshold $b_{\rm tresh}\sim \ln 2$, while, the gap between $E_{\rm deg}$ and $E_0$ still approximates the optimal value
\begin{align}
\label{eqapp:a_vs_deltaE}
    E_{\rm deg}-E_0=2b(N-1)-2a\sim \ln 2\;.
\end{align}
The exact finite value of $b$ is not important in this case, and can be taken as given. It follows that in such configurations, while $b\propto\mathcal{O}(1)$, while $a\sim (N-1)(b-\ln 2/2)$ scales linearly. Consequently we can obtain the error that is admitted on $a$ in these configurations. It follows, given Eq.~\eqref{eqapp:a_vs_deltaE} and the fact that the optimal gap has a fixed bandwidth tolerance $\mathcal{O}(1)$ (cf.~\ref{subsec:noise_degmod}), that a similar noise scaling applies to $a$, i.e.
\begin{align}
    a\sim \mathcal{O}(N)\;, \quad \delta a\sim \mathcal{O}(1)\;, \quad \frac{\delta a}{a}\sim \mathcal{O}(N^{-1})\;.
\end{align}

\paragraph{Subtler sources of parameter-noise.}

Finally, notice that in the implementation of $H_{\rm Star}$ a more general error could arise. That is, the actual tuning of the parameters of the generic spin Hamiltonian~\eqref{eq:classical_H_generic}
\begin{align}
     H=\sum_i^N h_i \sigma^{z}_i + \sum_{i < j}^N J_{ij} \sigma^{z}_i \sigma^{z}_j
\end{align}
to be converted into $H_{\rm Star}$ assumes all $J_{ij}=0$ for $i>1$ and $j>1$. Moreover, assuming (realistically) these contributions to be null, the resulting Hamiltonian is
\begin{align}
    H_{\rm Star-noisy}(a,\vec{b}^{(1)},\vec{b}^{(2)})=a\sigma^z_1+\sum_{i=2}^N b_i^{(1)}\sigma_i^z + \sum_{i=2}^N b_i^{(2)}\sigma_i^z\sigma_1^z\;.
\end{align}
The Star model assumes $b^{(1)}_i=b^{(2)}_i:=b$. Noise in the couplings might however affect this constraint. The main consequence would be a splitting of the level $E_{\rm deg}$ due to the fact that the configurations with $\sigma^z_1=-1$ would have a binomial spectrum
\begin{align}
  E_{\rm deg-noisy}=-a+\sum_{i=2}^N \sigma_i^z (b_i^{(1)}-b_i^{(2)})\;.  
\end{align}
The bandwidth splitting of $E_{\rm deg}$ is therefore characterized by
\begin{align}
    \big| \sum_{i=2}^N  b_i^{(1)}-b_i^{(2)} \big| \lesssim
\mathcal{O}(1)
\end{align}
where the allowed constant bandwidth was derived above~\ref{subsec:noise_degmod}.
It follows that, in general the error of each spin $i\geq 2$ should be of order $\mathcal{O}(N^{-1})$, i.e.
\begin{align}
    b^{(-)}_i:=|b_i^{(1)}-b_i^{(2)}| \lesssim \mathcal{O}(N^{-1})\;.
\end{align}
For what concerns 
\begin{align}
    b^{(+)}_i:=|b_i^{(1)}+b_i^{(2)}|= \mathcal{O}(N)\;
\end{align}
its scaling and relative error tolerance are the same as $b$, that is~\eqref{eqapp:ab_tol} for the optimal degeneracy case~\ref{par:noise_optimal_conf}, or ``irrelevant'' for the suboptimal configurations discussed above in~\ref{par:noise_suboptimal_conf}.

\section{Analytics for the Star model and Star-chain model}
In this Appendix we provide additional analytics regarding the two main models presented in the main text, i.e. the Star model~\eqref{eq:H_star1} and the Star-chain~\eqref{eq:starchain}.

\subsection{Partition Function for the Star model}
\label{app:analytics_star}

Given the energies and degeneracies indicated in Sec.~\ref{sec:star_model}, we can exactly compute the partition function $Z = \Tr[e^{-\beta H}]$ for the Star model~\eqref{eq:H_star1},
\begin{align}
    H_{\text{Star}[N]}(a,b) &:= a \,\sigma_1^z + b \sum_{i=2}^N \sigma_i^z  \left(\mathbb{1} + \sigma_1^z\right)  \;,
\end{align}
The partition function $Z=\sum_i e^{-\beta E_i}$ is
\begin{equation}
	Z_\text{Star}= \sum_{\vec{\sigma}^z}e^{-\beta H_{\rm Star}[\vec{\sigma^z}]} = e^{-\beta a}(e^{2\beta b}+ e^{-2\beta b})^{N-1} + 2^{N-1}e^{\beta a},
	\label{eqapp:Z_star}
\end{equation}
where the first term correspond to the binomial part of the spectrum (i.e. for $\sigma_1^z=1$), while the second term correspond to the $2^{N-1}$ degeneracy that is obtained for $\sigma_1^z=-1$.
The above expression can be manipulated into
\begin{equation}
\label{eqapp:zstar_cosh}
	Z_\text{Star}= 2^{N-1}\left(e^{-\beta a} \cosh(2\beta b)^{N-1}+e^{\beta a}\right)\;,
\end{equation}
and can be used to compute efficiently relevant quantities such as the average energy, the free energy etc., as from standard statistical mechanics. In particular the average energy is given by
\begin{align}
\langle H\rangle_\beta = \frac{\sum_{\vec{\sigma}^z} H[\vec{\sigma^z}] e^{-\beta H[\vec{\sigma^z}]}}{Z}=
-\frac{\partial}{\partial\beta}\ln Z\;.
\end{align}
Similarly the heat capacity, or energy variance, is given by
\begin{align}
     \Delta_\beta^2 H =\frac{\partial^2}{\partial\beta^2}\ln Z\;,
\end{align}
which for the Star model can be expressed analytically by substituting~\eqref{eqapp:Z_star}
\begin{align}
    \frac{
    4b^2 (N-1)\cosh(2 b)^N+ 2e^{2 a}\cosh(2 b)
    \left(a^2-b^2(N-1)(N-3)+(a^2+b^2(N-1)^2)\cosh(4 b)-2ab(N-1)\sinh(4 b)\right)}{\cosh(2 b)^{2-N}\left(e^{2 a}\cosh(2 b)+\cosh(2 b)^N\right)^2} 
\end{align}
in temperature units where $\beta=1$.

\subsection{Statistics of the energy levels, exponential suppression above the degeneracy, and slightly suboptimal configurations with better parameter scaling}
\label{subsec:star_statistics}

In this Section we analyze the statistics of the energy levels of the Star model. As we argued in App.~\ref{subsecapp:reduction}, the Star-chain model becomes equivalent to the former in the limit of large $|J|$.

The probability of a given energy outcome in from a Gibbs state is, in temperature units $\beta=1$, given by
\begin{align}
    P(E)=\frac{\sum_i e^{-E_i}\delta(E-E_i)}{\sum_j e^{-E_j}}=\frac{\sum_i e^{-E_i}\delta(E-E_i)}{Z}\;,
\end{align}
$Z$ being the partition function, which for the Star model is, from \eqref{eqapp:zstar_cosh}, in temperature units $\beta=1$,
\begin{align}
 Z_{\rm Star}=2^{N-1}\left(e^{-a} \cosh(2 b)^{N-1}+e^{a}\right)\;.
 \end{align}
Without loss of generality, it is possible to shift all energies in order to have $E_{\rm deg}$=0 for simplicity. Given that $E_{\rm deg}=-a$, this is equivalent to multiplying the partition function with a factor $e^{-a}$. That is, 
in this case the spectrum reduces to
\begin{align}
    E_0 =& 2a-2b(N-1)\;, &  \text{degeneracy }&1\;, \\
    E_{\rm deg}=& 0\;, &  \text{degeneracy }&2^{N-1}\;, \\
    E_k=&2a-2b(N-1-2k)\;, &  \text{degeneracy }&\binom{N-1}{k}\;.
\end{align}
and the partition function becomes
\begin{align}
    Z'=2^{N-1}\left(e^{-2a} \cosh(2 b)^{N-1}+1\right)
    =e^{2b(N-1)-2a} \left(1+e^{-4b}\right)^{N-1}+2^{N-1}\;.
\end{align}
Notice that it is possible to identify three contributions to $Z'$, i.e.
\begin{align}
    Z'=Z'P(E_0)+Z'P(E_{\rm deg})+Z'\sum_{k\geq 1}P(E_k)
\end{align}
corresponding respectively to the weight of the ground state, degenerate level, and all the binomial levels above $E_0$, i.e.
\begin{align}
    Z'P(E_0)=& e^{2b(N-1)-2a}\\
    Z'P(E_{\rm deg})=& 2^{N-1}\\
    Z'\sum_{k\geq 1}P(E_k)=& e^{2b(N-1)-2a} \left((1+e^{-4b})^{N-1}-1\right)\;.
    \label{eqapp:k_contr}
\end{align}
We now prove that in the optimal configurations, all the statistics of the Star model resides in $P(E_0)$ and $P(E_{\rm deg})$, while the remaining $\sum_{k\geq 1}P(E_k)$ is exponentially suppressed, as far as $b$ grows (at least) logaritmically. Moreover, in the optimal configurations, we know from the main text and from App.~\eqref{app:analytics_star}
that $2b(N-1)-2a\sim (N-1)\ln 2$, and therefore in such case one has
$P(E_0)\sim P(E_{\rm deg)}\sim \frac{1}{2}$.
Finally, as far as $b$ grows faster than $\ln N$, all the statistical contribution from the other levels $E_{k\geq 1}$ is suppressed. This can be seen from Eq.~\eqref{eqapp:k_contr} and
\begin{align}
    (1+e^{-4b})^{N-1}= \left(1+\frac{e^{-({4b}-{\ln (N-1)})}}{N}\right)^{N-1}\;,
\end{align}
which, for large $N$, tends to
\begin{align}
    \left(1+\frac{e^{-({4b}-{\ln (N-1)})}}{N}\right)^{N-1} \rightarrow e^{e^{-({4b}-{\ln (N-1)})}}\;.
\end{align}
For example, in the optimal configuration of the Star model, $4b\sim (N-1)\ln 2$ grows linearly in $N$ and the whole contribution of the statistics from all levels $E_{k\geq 1}$, is suppressed exponentially as
\begin{align}
    e^{e^{-({4b}-{\ln (N-1)})}} -1 \rightarrow e^{-2^{N-1}} - 1 \sim \frac{1}{2^{N-1}}.
\end{align}

\subsection{Star-chain: partition function and spectrum}
\label{app:analytics_starchain}
\paragraph{Partition function of the Star-chain.}

Remarkably, the Star-chain model at equilibrium can be exactly solved, in the sense that it is possible to compute analytically its partition function, using the \emph{transfer matrix method}, which is used in standard solutions of the 1D Ising model~\cite{onsager1944ising}.
Consider the Star-chain Hamiltonian
\begin{align}
 H_\text{Star-chain}=
 a \sum_\alpha \sigma^z_\alpha + J \sum_\alpha \sigma^z_\alpha \sigma^z_{\alpha+1} + b \sum_{\alpha,i} (\sigma^z_\alpha+\mathbb{1}) \sigma^z_{\alpha,i}\;, \quad \alpha=1,\dots,n\;, \ \ i=1,\dots,m\;.
\end{align}
The partition function is given by definition as
\begin{align}
    	Z_\text{Star-chain}= \sum_{\vec{\sigma}^z}e^{-\beta H_\text{Star-chain}[\vec{\sigma}^z]}, 
\end{align}
where $\vec{\sigma}^z$ is the $(n+nm)$-long vector given by $\vec{\sigma}^z=\{\sigma^z_\alpha,\sigma^z_{\alpha,i}\}$ and is summed over all possible values $\pm 1$ of all the spins. It is therefore possible to separate the two classes of spins by defining 
\begin{align}
    \vec{\sigma}^z_{(1)}=\{\sigma^z_\alpha\}\;, \quad
    \vec{\sigma}^z_{(2)}=\{\sigma^z_{\alpha,i}\}\;.
\end{align}
To compute the partition function we can consider
\begin{align}
\label{eqapp:Zdoublesum}
    	Z_\text{Star-chain}= \sum_{\vec{\sigma}^z_{(1)}}\sum_{\vec{\sigma}^z_{(2)}}e^{-\beta H_\text{Star-chain}[\vec{\sigma}^z]}\;.
\end{align}
Moreover 
for fixed $\vec{\sigma}^z_{(1)}=\{\sigma^z_\alpha\}$, we can re-express the second sum as
\begin{align}
\label{eqapp:Wappears}
   \sum_{\vec{\sigma}^z_{(2)}}e^{-\beta H_\text{Star-chain}[\vec{\sigma}^z]}=
   \prod_{\alpha}\sum_{\{\sigma^z_{\alpha,i}\}} e^{-\beta a\sigma^z_{\alpha}}e^{-\beta J \sigma^z_{\alpha}\sigma^z_{\alpha+1}} e^{-\beta b\sum_i (\sigma^z_\alpha+\mathbb{1}) \sigma^z_{\alpha,i}} := \prod_{\alpha} W(\sigma^z_\alpha,\sigma^z_{\alpha+1})\;,\\
   \text{with }\quad  W(\sigma^z_\alpha,\sigma^z_{\alpha+1})\equiv \sum_{\{\sigma^z_{\alpha,i}\}} e^{-\beta a\sigma^z_{\alpha}}e^{-\beta J \sigma^z_{\alpha}\sigma^z_{\alpha+1}} e^{-\beta b\sum_i (\sigma^z_\alpha+\mathbb{1}) \sigma^z_{\alpha,i}}\;.
\end{align}
We now notice that, when fixing $\vec{\sigma}^z_{(1)}=\{\sigma^z_\alpha\}$ it is possible to solve the sum over $\vec{\sigma}^z_{(2)}=\{\sigma^z_{\alpha,i}\}$ by using the fact that
\begin{align}
    \sigma^z_\alpha=1 &\Rightarrow \sum_{\sigma^z_{\alpha,i}}e^{-\beta(\sigma^z_\alpha+1)\sum_i\sigma^z_{\alpha,i}}=\prod_i \sum_{\sigma^z_{\alpha,i}}e^{-\beta(\sigma^z_\alpha+1)\sigma^z_{\alpha,i}}=\left(e^{2\beta b}+e^{-2\beta b}\right)^m\;.\\
    \sigma^z_\alpha=-1 &\Rightarrow \sum_{\sigma^z_{\alpha,i}}e^{-\beta(\sigma^z_\alpha+1)\sum_i\sigma^z_{\alpha,i}}=\prod_i \sum_{\sigma^z_{\alpha,i}}e^{-\beta(\sigma^z_\alpha+1)\sigma^z_{\alpha,i}}=2^m\;.
\end{align}
It follows that $W(\sigma^z_{\alpha},\sigma^z_{\alpha+1})$ can be seen as a $2\times 2$ matrix (corresponding to the four elements $(\sigma^z_{\alpha},\sigma^z_{\alpha+1})=(\pm 1, \pm 1)$)\;,
\begin{align}
  W(\sigma^z_{\alpha},\sigma^z_{\alpha+1})=
  \begin{pmatrix}
  e^{-\beta a} e^{-\beta J} \left(e^{2\beta b}+e^{-2\beta b}\right)^m 
  & e^{-\beta a} e^{\beta J} \left(e^{2\beta b}+e^{-2\beta b}\right)^m \\
  e^{\beta a} e^{\beta J} 2^m
  & e^{\beta a} e^{-\beta J} 2^m
  \end{pmatrix}\;.
\end{align}
Finally, notice that the partition function is given by (cf. \eqref{eqapp:Zdoublesum} and \eqref{eqapp:Wappears})
\begin{align}
\label{eqapp:ZSC_lam}
    Z_\text{Star-chain}=\sum_{\{\sigma^z_\alpha\}}\prod W(\sigma^z_\alpha,\sigma^z_{\alpha+1})=\Tr[W^n]=\lambda_+^n+\lambda_-^n\;,
\end{align}
where $\lambda_+$ and $\lambda_-$ are the two eigenvectors of $W$, that are, in temperature units $\beta=1$,
\begin{align}
    \lambda_+= & 2^{m-1}e^{-a-J}
    \left( e^{2J}(e^{2a}+\cosh{(2b)}^m)
    +\sqrt{4e^{2a}\cosh{(2b)}^m+ e^{4J}(e^{2a}-\cosh{(2b)}^m)^2}
    \right)\;,\\
    \lambda_-= & 2^{m-1}e^{-a-J}
    \left( e^{2J}(e^{2a}+\cosh{(2b)}^m)
    -\sqrt{4e^{2a}\cosh{(2b)}^m+ e^{4J}(e^{2a}-\cosh{(2b)}^m)^2}
    \right)\;.
\end{align}
Substituting these values in~\eqref{eqapp:ZSC_lam} constitutes the analytical expression of the partition function for the Star-chain model.

\paragraph{Spectrum of the Star-chain.}
In this subsection we will build analytical considerations on the energy spectrum of the Star-chain model.
Consider again the Hamiltonian 
\begin{align}
 H_\text{Star-chain}=
 a \sum_\alpha \sigma^z_\alpha + J \sum_\alpha \sigma^z_\alpha \sigma^z_{\alpha+1} + b_1 \sum_{\alpha,i} \sigma^z_\alpha \sigma^z_{\alpha,i} + b_2 \sum_{\alpha,i} \sigma^z_{\alpha,i}\;.
\end{align}
Here $\alpha=1,...,n$ indices the privileged spins, and $i=1,...,m$ the ``subordinate spins" of each $\alpha$-spin, for a total of $N=n(m+1)$ spins. Also, we can take the ``Star-choice" $b_1=b_2$ that guarantees degeneracy. We are left with
\begin{align}
 J\mathcal{I}(\sigma^z_\alpha)+ a \sum_\alpha \sigma^z_\alpha + b \sum_{\alpha,i} (\sigma^z_\alpha+\mathbb{1}) \sigma^z_{\alpha,i} \qquad \text{with } \mathcal{I}(\sigma_\alpha^z):=\sum_\alpha \sigma_\alpha^z \sigma_{\alpha+1}^z\;.
\end{align}
We separated the term $\mathcal{I}(\sigma_\alpha^z)$ as it is the one that breaks the permutation symmetry (for cyclic boundary conditions it has only cyclic symmetry). Such term needs therefore all its $2^n$ levels to be resolved.
The remaining part is permutationally symmetric and therefore has a spectrum that can be computed efficiently. 
Define
\begin{align}
    n_\uparrow + n_\downarrow =n \qquad n_\uparrow:=\{\# \ \alpha-\text{spins up}\}
\end{align}
It follows that
\begin{align}
    a \sum_\alpha \sigma^z_\alpha = a (2n_\uparrow -n)\;.
\end{align}
Then, for the remaining spins, notice that there are $n_\downarrow m=(n-n_\uparrow)m$ that don't contribute to the energy due to their $\alpha$-spin being down, and the interaction of the form $\sum_{\alpha,i} (\sigma^z_\alpha+\mathbb{1}) \sigma^z_{\alpha,i}$. For the same reason, all the remaining one see an effective magnetic field equal to $2b$. We therefore define
\begin{align}
    \mu_\uparrow+\mu_\downarrow =m n_\uparrow \qquad \mu_{\uparrow}:=\{\# \text{subordinate spins up with their $\alpha$-spin up}\}
\end{align}
It follows that
\begin{align}
    b \sum_{\alpha,i} (\sigma^z_\alpha+\mathbb{1}) \sigma^z_{\alpha,i} = 2b (2\mu_\uparrow - mn_\uparrow)\;.
\end{align}
Putting all the pieces together, we cannot coarse grain easily the $2^n$ degeneracy of
the $\sigma_\alpha^z$ configurations, but we can simplify the remaining degeneracy by writing down the energy levels as
\begin{align}
    E= J\mathcal{I}(\sigma_\alpha^z) + a (2n_\uparrow -n) + 2b (2\mu_\uparrow - mn_\uparrow)
\end{align}
with $n_\uparrow=1,\dots,n$ fixed by the configuration $\sigma_\alpha^z$, $\mu_\uparrow=1,...,mn_\uparrow$, and degeneracy (for each $\sigma_\alpha^z$-configuration) equal to
\begin{align}
    2^{(n-n_\uparrow)m}\binom{m n_\uparrow}{\mu_\uparrow}\;.
\end{align}

\section{All-to-All model}
\label{app:alltoall_model}
 
Consider the following model of an $N$-spin Hamiltonian.
\begin{equation}
	H_\text{All}(h, J) := -h \sum_{i=1}^N \sigma_i^z   -J \sum_{i<j}^N \sigma_i^z  \sigma_j^z,
	\label{eqapp:alltoall}
\end{equation}
where $h$ and $J$ are two coefficients. This model consistently emerged from numerical optimisation of $\mathcal{C}$ for small number of spins, up to $N=5$.
Moreover, the Hamiltonian~\eqref{eqapp:alltoall} model is completely symmetric under permutations of the spins' operators. This helps in expressing its spectrum as a function of the total number $k$ of spins up, having $\sigma^z_i=+1$ (it follows that $N-k$ spins are in the opposite configuration, $\sigma_i^z=-1$)
\begin{equation}
	E_k = h(N-2k) +\frac{J}{2}\left[ 4k(N-k) - N(N-1)  \right],
	\label{eqapp:a2a_Ek}
\end{equation}
each level with degeneracy
\begin{equation}
    \text{deg}[E_k]=\binom{N}{k}\;.
\end{equation}
It follows that the partition function is given by
\begin{equation}
	Z_\text{all} = \sum_{k=0}^N \binom{N}{k} e^{-\beta E_k}.
\end{equation}
From numerical optimization (cf. App.~\ref{appsub:adam_at_various_N}), it appears that the optimal values of $h$ and $J$ that maximise $\C$ in this model satisfy the relation
\begin{equation}
	h=J.
\end{equation}
Under such hypothesis, the above expression~\eqref{eqapp:a2a_Ek} for the energy levels can be written as 
\begin{equation}
	\frac{E_k}{J} = -\frac{N(N+1)}{2}+2(k+1)(N-k)=E_{k=N}+2(k+1)(N-k)\;,
\end{equation}
which clarifies explicitly the ground state being $E_{k=N}$ and the fact that all the levels above form a spectrum that is parabolic in $k$, with a first excited level corresponding to $k=N-1$ and $k=0$, with total degeneracy $\text{deg}[E_{k=0}]+\text{deg}[E_{k=N-1}]=N+1$.

\end{document}